\documentclass[12pt]{article}%
\usepackage[top=3cm, bottom=3cm, left=3cm, right=3cm]{geometry}
\usepackage{amsmath}
\usepackage{amsfonts}
\usepackage{amssymb}
\usepackage{graphicx}%

\usepackage{cite}

\usepackage{xcolor}

\long\def\/*#1*/{}

\providecommand{\U}[1]{\protect\rule{.1in}{.1in}}
\begin{document}

\title{QFT treatment of a bound state in a thermal gas
}
\author{Subhasis Samanta$^{1}$
\and  Francesco Giacosa$^{1,2}$ \\$^{1}$\textit{Institute of Physics, Jan-Kochanowski University, }\\\textit{ul. Swietokrzyska 15, 25-406, Kielce, Poland.}\\$^{2}$\textit{Institute for Theoretical Physics, J. W. Goethe University, }\\\textit{ Max-von-Laue-Str. 1, 60438 Frankfurt, Germany.}}
\date{}
\maketitle

\begin{abstract}

We investigate how to include bound states in a thermal gas
in the context of quantum field theory (QFT). To this end, we use for
definiteness a scalar QFT with a $\varphi^{4}$ interaction, where the field $\varphi$ represents a particle with mass $m$. A bound
state of the $\varphi$-$\varphi$ type is created when the coupling constant is negative
and its modulus is
larger than a certain critical value. We investigate the contribution of this
bound state to the pressure of the thermal gas of the system by using the
$S$-matrix formalism involving the derivative of the phase-shift scattering.
Our analysis, which is based on an unitarized one-loop resumed approach which
renders the theory finite and well-defined for each value of the coupling
constant, leads to following main results: (i) We generalize the phase-shift
formula in order to take into account within a unique formal approach the
two-particle interaction as well as the bound state (if existent). (ii)
\textit{On the one hand}, the number density of the bound state in the system at a certain
temperature $T$ is obtained by the standard thermal integral; this is the case for any binding energy, even if it is much
smaller than the temperature of the thermal gas. (iii) \textit{On the other
hand}, the contribution of the bound state to the total pressure is partly
-but not completely- canceled by the two-particle interaction contribution to the pressure.
(iv) The pressure as function of the coupling constant is \textit{continuous}
also at the critical coupling for the bound state formation: the jump in
pressure due to the sudden appearance of the bound state is exactly canceled 
by an analogous jump (but
with opposite sign) of the interaction contribution to the pressure.

\end{abstract}

\section{Introduction}

Measurement of bound states, such as deuteron ($d$), helium-3 ($^{3}$He), tritium ($^{3}\text{H}$), helium-4 ($^{4}\text{He}$), hypertritium
($_{\Lambda}^{3}$H) and their antiparticles, was reported in high energy pp,
proton-nucleus (pA) and nucleus-nucleus (AA) collisions \cite{Cocconi:1960zz,
Abelev:2010rv,Agakishiev:2011ib,Adam:2015vda,Adam:2019phl,Acharya:2019xmu,Acharya:2020sfy}%
. Moreover, the QCD spectrum has also revealed the existence of a whole new
class of $X,$ $Y,$ and $Z$ resonances that are not predicted by the quark
model, some of which can be mesonic molecular bound states, see e.g. Ref.
\cite{Esposito:2014rxa} and references therein.  Last but not least, also pentaquark states \cite{Aaij:2019vzc} can be understood as molecular objects.

The production of nuclei as well as other hadronic bound states has attracted a lot of interest because their binding energies are typically much smaller than the temperature realized in high energy collisions, hence at the first sight it is quite puzzling that such objects can form in such a hot environment.  
In addition, light nuclei are also potential candidates to
search for the critical point in the quantum chromodynamics (QCD) phase
diagram \cite{Sun:2017xrx,Sun:2018jhg,Liu:2019nii,Luo:2020pef}. Excess
production of some light anti-nuclei in cosmic rays and dark matter
experiments \cite{Blum:2017qnn,Poulin:2018wzu,Vagelli:2019tqy} has also
been investigated. 

There are several models, notably thermal models
\cite{Siemens:1979dz,Andronic:2010qu,Andronic:2012dm,Cleymans:2011pe,Ortega:2017hpw,
Ortega:2019fme}, nucleon coalescence models
\cite{Gutbrod:1988gt,Sato:1981ez,Mrowczynski:1992gc,Csernai:1986qf,
Mrowczynski:2016xqm,Bazak:2018hgl,Dong:2018cye,Sun:2016rev,
Sun:2018jhg,Polleri:1997bp, Mrowczynski:2019yrr, Bazak:2020wjn}, and dynamical
models \cite{Danielewicz:1991dh,Oliinychenko:2018ugs} which aim to explain the
production of bound states in high energy collisions.
Yet, there are differences among them, and it is not yet clear up to now which
approach is the correct one. In other words, are bound states produced according to their statistic distribution at temperature $T?$ If yes,
which is their contribution to the pressure?

In the present work, we intend to answer these questions in the context of Quantum Field Theory (QFT).
To this end, we use the well known scalar $\varphi^{4}$-interaction, where
$\varphi$ is a field with mass $m.$

First, we evaluate the scattering phase-shift at tree-level and at
the one-loop resumed level. In the latter (and necessary) step, we choose a
proper unitarization scheme at the resumed one-loop level for which (i) no
new energy scale appears and (ii) the results are finite and well-defined for
any value of the coupling constant, denoted as $\lambda$ 
(the corresponding potential reads $V=\lambda ^{4}\varphi ^{4}/4!$).

When $\lambda>0$ the interaction is repulsive, and the phase-shift is always
decreasing with the increase of the running energy $\sqrt{s}$
and smaller than zero. When $\lambda<0$ (and its modulus is smaller
than a certain critical value denoted as $\lambda _{c}$) the interaction is attractive and the
phase-shift is positive, rising for small $\sqrt{s}$ and decreasing afterward.
Yet, when $\lambda<\lambda_{c}<0,$ a bound-state is formed, whose mass is
exactly equals to $2m$ for $\lambda=\lambda_{c}$ and is smaller than $2m$ for
$\lambda<\lambda_{c}.$ In this case, the interaction is again repulsive and
the phase-shift is negative and decreasing.

We use the previous results to study the properties of this QFT at finite
temperature by using the phase-shift (or S-matrix) approach, according to
which the density of states is proportional to the derivative of the
phase-shift w.r.t. the $\sqrt{s}$. For $\lambda>0$, the
contribution of the interaction to the pressure (as well as to other
quantities) is negative, in agreement with the repulsive nature of the
interaction. On the other hand, for $\lambda_{c}<\lambda<0,$ the contribution
to the pressure is positive, as the attraction suggests.

The case $\lambda<\lambda_{c}$ requires care: on the one hand, the repulsion
causes a negative contribution of the $\varphi$-$\varphi$ interaction to the pressure, but the presence of the
bound state implies a positive contribution to the pressure: the net result is
a positive contribution. Quite remarkably, the total pressure as function of the
coupling constant $\lambda$ is \textit{continuous} also at $\lambda=\lambda_{c}$: the jump in
pressure generated by the abrupt appearance of the bound state is
\textit{exactly} canceled by an analogous jump (but with opposite sign) due
to the phase-shift contribution to the pressure. 
Within this context, we shall extend the S-matrix formalism to
include the contribution of eventual bound states. 
This point represents a formal achievement of our
approach and corresponds to a rather intuitive aspect of the problem: the
bound state is also an outcome of the two-particle interaction, hence its role
should be also described by a (proper) extension of the phase-shift approach below the particle-particle threshold.

In summary, our findings show that the number density
of the bound state with mass $M_B$ can be calculated by the
`simple' thermal integral%
\begin{equation}
n_{B}\overset{\text{ }}{=}\theta(\lambda_{c}-\lambda)\int_{k}\left[
e^{-\beta\sqrt{k^{2}+M_{B}^{2}}}-1\right]  ^{-1}%
\end{equation}
for any temperature $T$ (in the previous equation, $\int_{k} \equiv \int d^3k/(2\pi)^3$). This result is valid also when the mass of the bound
state $M_{B}$ is just below the threshold $2m$ and for temperatures
$T\gg2m-M_{B}$ (hence, even for temperatures much larger than the binding energy). However, the
contribution of the interacting $\varphi\varphi$-system is \textbf{not }simply
given by the standard contribution to the pressure%
\begin{equation}
P_{B}\overset{\text{ }}{=}-\theta(\lambda_{c}-\lambda)T\int_{k}\ln\left[
1-e^{-\beta\sqrt{k^{2}+M_{B}^{2}}}\right]  \text{ }, \label{pbint}%
\end{equation}
but caution is needed.  In general, we shall find that for $\lambda<\lambda_{c}$ the total
interacting contribution to the pressure (including both the bound state and the $\varphi\varphi$-interaction above threshold) 
can be expressed as
\begin{equation}
\zeta P_{B}\text{ with }0<\zeta<1.
\end{equation}
For small temperatures, the ratio $\zeta$ is close to $1$, but for higher temperatures it
saturates to a certain finite which is typically about $0.5$. Quite
interestingly, the existence of this cancellation was discussed in the framework of
Quantum\ Mechanics (QM) in Ref, \cite{Ortega:2017hpw}, even if in that
case the cancellation was more pronounced ($\zeta$ quite small) than the result obtained in our QFT approach.
 
In conclusion, when a bound state forms in a thermal gas, one should not
simply add the corresponding thermal integral
as in\ Eq. (\ref{pbint}) to the pressure, since the additional role of the interaction that leads to the very existence of that bound state is not negligible and contributes with an opposite sign.

The paper is organized as follows: in Sec. 2 we concentrate on the main properties of the system in the vacuum, that include phase-shifts, unitarization procedure, and the emergence of a bound state when the attraction is strong enough; then, in Sec. 3  we present the results at nonzero temperature with special focus on the pressure and the role of the bound state; finally, in Sec. 4 we summarize and conclude our paper.


\section{Vacuum phenomenology of scalar $\varphi^{4}$-theory}

\subsection{Scattering phase shifts}

\label{sec:scattering} In this section we discuss the relatively simple but
nontrivial interacting QFT involving a single scalar field $\varphi$ subject to the Lagrangian%

\begin{equation}
\mathcal{L}=\frac{1}{2}\left(  \partial_{\mu}\varphi\right)  ^{2}-\frac{1}%
{2}m^{2}\varphi^{2}-\frac{\lambda}{4!}\varphi^{4}, \label{eq:L}%
\end{equation}
where the first two terms describe a free particle with mass $m$ and the last
term corresponds to the quartic interaction. The coupling constant $\lambda$ is
dimensionless and the theory is renormalizable \cite{Peskin:1995ev}. For a
detailed analysis of this theory in the context of perturbation
theory\footnote{The $\varphi^{4}$ QFT could also be trivial, in the sense that
the coupling constant vanishes after the renormalization procedure is carried out, see e.g.
Refs. \cite{Wolff:2009ke, Agodi:1997qt} and refs. therein for the discussion of
this issue.} see Ref. \cite{Kleinert:2001ax}. As we shall comment later on, we
will introduce a non-perturbative unitarization procedure on top of Eq.
(\ref{eq:L}), in such a way to make the theory finite, unitary and well defined for each value of the coupling constant $\lambda$
(even for large ones). This is done at the one-loop resummed level with a
suitable subtraction constant.

In the centre of mass frame, the differential cross-section is given by \cite{Peskin:1995ev}
\begin{equation}
\frac{d\sigma}{d\Omega}=\frac{|A(s,t,u)|^{2}}{64\pi^{2}s},
\end{equation}
where $A(s,t,u)$ is the scattering amplitude as evaluated through Feynman
diagrams, and $s,t$ and $u$ are Mandelstam variables:
\begin{align}
s  &  =(p_{1}+p_{2})^{2}\geqslant4m^{2}\text{ ,}\\
t  &  =(p_{1}-p_{3})^{2}=-\frac{1}{2}(s-4m^{2})(1-\cos\theta)\leq0\text{ ,}\\
u  &  =(p_{2}-p_{3})^{2}=-\frac{1}{2}(s-4m^{2})(1+\cos\theta)\leq0\text{ ,}%
\end{align}
where $p_{1},p_{2},p_{3}$ and $p_{4}$ are four-momenta of the particles
($p_{1},p_{2}$ ingoing and $p_{3},p_{4}$ outgoing), and $\theta$ is the
scattering angle. The sum of these three variables is $s+t+u=4m^{2}$. The
scattering amplitude can be expressed in terms of partial waves (by keeping
$s$ and $\theta$ as independent variables) as \cite{Messiah}:%

\begin{equation}
A(s,t,u)=A(s,\theta)=\sum_{l=0}^{\infty}(2l+1)A_{l}(s)P_{l}(\cos
\theta)\text{{ ,}}%
\end{equation}
where $P_{l}(\xi)$ with $\xi=\cos\theta$ are the Legendre polynomials with
\begin{equation}
\int_{-1}^{+1}d\xi P_{l}(\xi)P_{l^{\prime}}(\xi)=\frac{2}{2l+1}\delta
_{ll^{\prime}}\text{ .}%
\end{equation}
In general, the $l$-th wave contribution to the amplitude is given by
$A_{l}(s)=\frac{1}{2}\int_{-1}^{+1}d\xi A(s,\theta)P_{l}(\xi).$

In the particular case of our Lagrangian of Eq. (\ref{eq:L}), the tree-level
scattering amplitude $A(s,t,u)$ takes the very simple form:
\begin{equation}
iA(s,t,u)=i(-\lambda)\Rightarrow A(s,t,u)=A(s,\theta)=-\lambda.
\end{equation}
For $\lambda>0$, one has $A<0$: the (tree-level) interaction is repulsive. On the other
hand for $\lambda<0$ one has $A>0,$ which corresponds to an attractive
interaction. [This case implies that the vacuum $\varphi=0$ is only metastable,
but this shall not affect our discussion.] 

At tree-level the $s$-wave amplitude's contribution takes the form:
\begin{equation}
A_{0}(s)=\frac{1}{2}\int_{-1}^{+1}d\xi A(s,\theta)=A(s,\theta)=-\lambda\text{
,}%
\end{equation}
while all other waves vanish, $A_{l=1,2,...}(s)=0$ (this
holds true also when unitarizing the theory
within the adopted resummation scheme). Further, the total
cross-section reads

\begin{equation}
\sigma(s)=\frac{1}{2}2\pi\frac{1}{64\pi^{2}s}\sum_{l=0}^{\infty}%
2(2l+1)\left\vert A_{l}(s)\right\vert ^{2}=\frac{1}{32\pi%
s}\left\vert A_{0}(s)\right\vert ^{2}. \label{eq:sigma}%
\end{equation}
At threshold:%
\begin{equation}
\sigma(s_{th}=4m^{2})=\frac{1}{2}2\pi\frac{1}{64\pi^{2}s}2\left\vert
A_{0}(s_{th})\right\vert ^{2}=8\pi\left\vert a_{0}^{\text{SL}}\right\vert
^{2},
\end{equation}
where $a_{0}^{\text{SL}}$ is the s-wave ($l=0$) scattering length (at
tree-level) given by:%

\begin{equation}
a_{0}^{\text{SL}}=\frac{1}{2}\frac{A_{0}(s=4m^{2})}{8\pi\sqrt{4m^{2}}}%
=\frac{1}{2}\frac{-\lambda}{16\pi m}\text{ .}%
\end{equation}
The factor $1/2$ in the previous equation refers to identical particles.

\begin{figure}[ptb]
\centering
\includegraphics[width=0.48\textwidth]{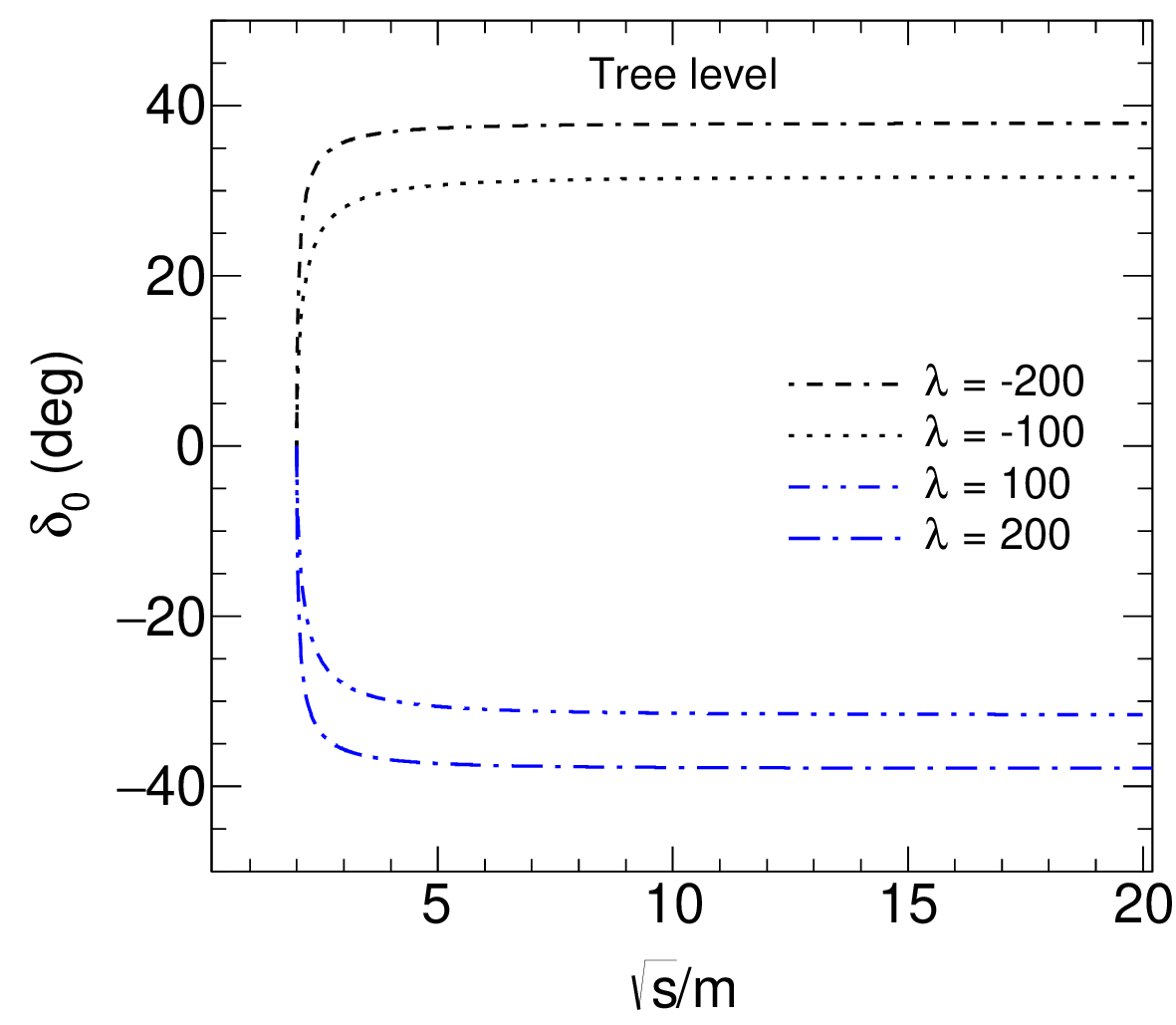}\caption{Behavior of the phase-shift at the tree-level for different values of
$\lambda$.}
\label{fig:phaseshift_treelevel}
\end{figure}

Next, we introduce the phase shifts. For identical particles, one has the
following general definition of the $l$-th wave phase shift $\delta_{l}(s)$:%
\begin{equation}
\frac{e^{2i\delta_{l}(s)}-1}{2i}=ka_{l}(s)=\frac{1}{2}\cdot\frac{k}{8\pi
\sqrt{s}}A_{l}(s) \text{ ,} \label{defdeltal}%
\end{equation}
where $k=\sqrt{\frac{s}{4}-m^{2}}$ is the modulus of the three-momentum of one of the ingoing (or outgoing) particles. In the present case, the only
non-vanishing phase-shift is given by $\delta_{0}(s)$%
\begin{equation}
\frac{e^{2i\delta_{0}(s)}-1}{2i}=ka_{0}(s)=\frac{1}{2}\cdot\frac{k}{8\pi
\sqrt{s}}A_{0}(s)\text{ ,} \label{defdelta0}%
\end{equation}
where the `running' length $a_{0}(s)$ is by construction such that
$a_{0}(s=4m^{2})=a_{0}^{\text{SL}}.$ Note, for $s$ just above the threshold we
have%
\begin{equation}
\frac{e^{2i\delta_{0}(s)}-1}{2i}\simeq\delta_{0}(s)\simeq ka_{0}^{\text{SL}}.
\end{equation}
In general, the phase-shift $\delta_{0}(s)$ can be calculated as:
\begin{equation}
\delta_{0}(s)=\frac{1}{2}\arg\left[  1-\frac{1}{16\pi}\sqrt{\frac{4m^{2}}{s}%
-1}A_{0}(s)\right]  \text{ .} \label{eq:ps_tree}%
\end{equation}%

Next, we explore the role of $\lambda$ for the tree-level scattering. In
Fig. \ref{fig:phaseshift_treelevel} we show the behavior of phase shift
$\delta_{0}(s)$ using Eq. (\ref{eq:ps_tree}) for different values of $\lambda$. 
For positive $\lambda$ values, the function $\delta
_{0}(s)$ is negative and decreases with increasing $\sqrt{s}/m$: the slope of
the curve ($\partial\delta_{0}/\partial\sqrt{s}$) is negative for any arbitrary
value of $s$, which indicates an repulsive interaction. For negative $\lambda$
values, the opposite behavior is realized, signalizing attraction.

The asymptotic values $\delta_{0}(s\rightarrow\infty)$ does not tend to a
multiple of $\pi,$ since the theory at first order in $\lambda$ is only
unitary at that order. As a consequence, we can trust the results only when
$\delta_{0}(s)$ is sufficiently small. As a related side-remark, the
expression $\delta_{0}(s)=\frac{1}{2}\arcsin\left[  \frac{k}{8\pi\sqrt{s}%
}A_{0}(s)\right]  $ (which in principle follows from Eq. (\ref{defdelta0})) is
also valid only when the amplitude is sufficiently small. This drawback is also due to the lack
of unitarity.

All these aspects show that the unitarization is necessary, as we show in
detail in the next subsection.

\subsection{Unitarization}

\label{sec:Unitarization} Here, we introduce the two-particle loop of the
field $\varphi$, that we denote as $\Sigma(s).$ We start from the requirement
about its imaginary part above threshold (because of the optical theorem):
\begin{equation}
I(s)=\operatorname{Im}\Sigma(s)=\frac{1}{2}\frac{\sqrt{\frac{s}{4}-m^{2}}%
}{8\pi\sqrt{s}}\text{for }\sqrt{s}>2m . \label{imat}
\end{equation}
We shall put here no cutoff, hence the above equation is considered valid up
to arbitrary values of the variable $s$ (note, in each realistic QFT the quantity
$\operatorname{Im}\Sigma(s)$ should decrease for $s$ large enough, e.g. above the GUT or the Planck scale; nevertheless, from a
mathematical point of view, we can get a fully consistent treatment for any value of $s$).
The loop function $\Sigma(s)$ for complex values of the variable $s$ reads
\begin{equation}
\Sigma(s)=\frac{1}{\pi}\int_{4m^{2}}^{\infty}ds^{\prime}\frac{I(s^{\prime}%
)}{s^{\prime}-s-i\epsilon}-C\text{ ,}
\end{equation}
where the subtraction $C$ guarantees convergence. Here, we make the choice
$\Sigma(s\rightarrow0)=0,$ hence
\begin{equation}
C=\frac{1}{\pi}\int_{4m^{2}}^{\infty}ds^{\prime}\frac{I(s^{\prime})}%
{s^{\prime}}.
\end{equation}
This choice turns out to be very convenient for our purposes. Explicitly, the
loop reads (we keep track of the arbitrary small $\epsilon$ since this will be
important later on):%

\begin{equation}
\Sigma(s)=\frac{1}{2}\frac{1}{16\pi}\left(  -\frac{1}{\pi}\sqrt{1-\frac
{4m^{2}}{s+i\epsilon}}\ln\frac{\sqrt{1-\frac{4m^{2}}{s+i\epsilon}}+1}%
{\sqrt{1-\frac{4m^{2}}{s+i\epsilon}}-1}\right)  +\frac{1}{16\pi^{2}}\text{.}
\label{eq:loop}%
\end{equation}
(For details on the $\varphi \varphi$ loops, see Ref. \cite{Giacosa:2007bn} and references therein.)
For $s$ being real we get
\begin{equation}
\operatorname{Im}\Sigma(s)=\left\{
\begin{array}
[c]{c}%
\frac{1}{2}\frac{\sqrt{\frac{s}{4}-m^{2}}}{8\pi\sqrt{s}}\text{for }s>\left(
2m\right)  ^{2}\\
\varepsilon\text{ for }s<\left(  2m\right)  ^{2},%
\end{array}
\right.  \label{imsigma}%
\end{equation}
where $\varepsilon\propto\epsilon$ is an infinitesimal positive quantity. Note. Eq. (\ref{imat}) is fulfilled, as it should.
Moreover, for $s$ real and larger than $4m^2$, the real part of the loop is given by the principal part (P) of the
following integral:
\begin{equation}
\operatorname{Re}\Sigma(s)=\frac{s}{\pi}P\int_{s_{th}}^{\infty}%
\frac{I(s^{\prime})}{(s^{\prime}-s)s^{^{\prime}}}.
\end{equation}

\begin{figure}[ptb]
\centering
\includegraphics[width=0.48\textwidth]{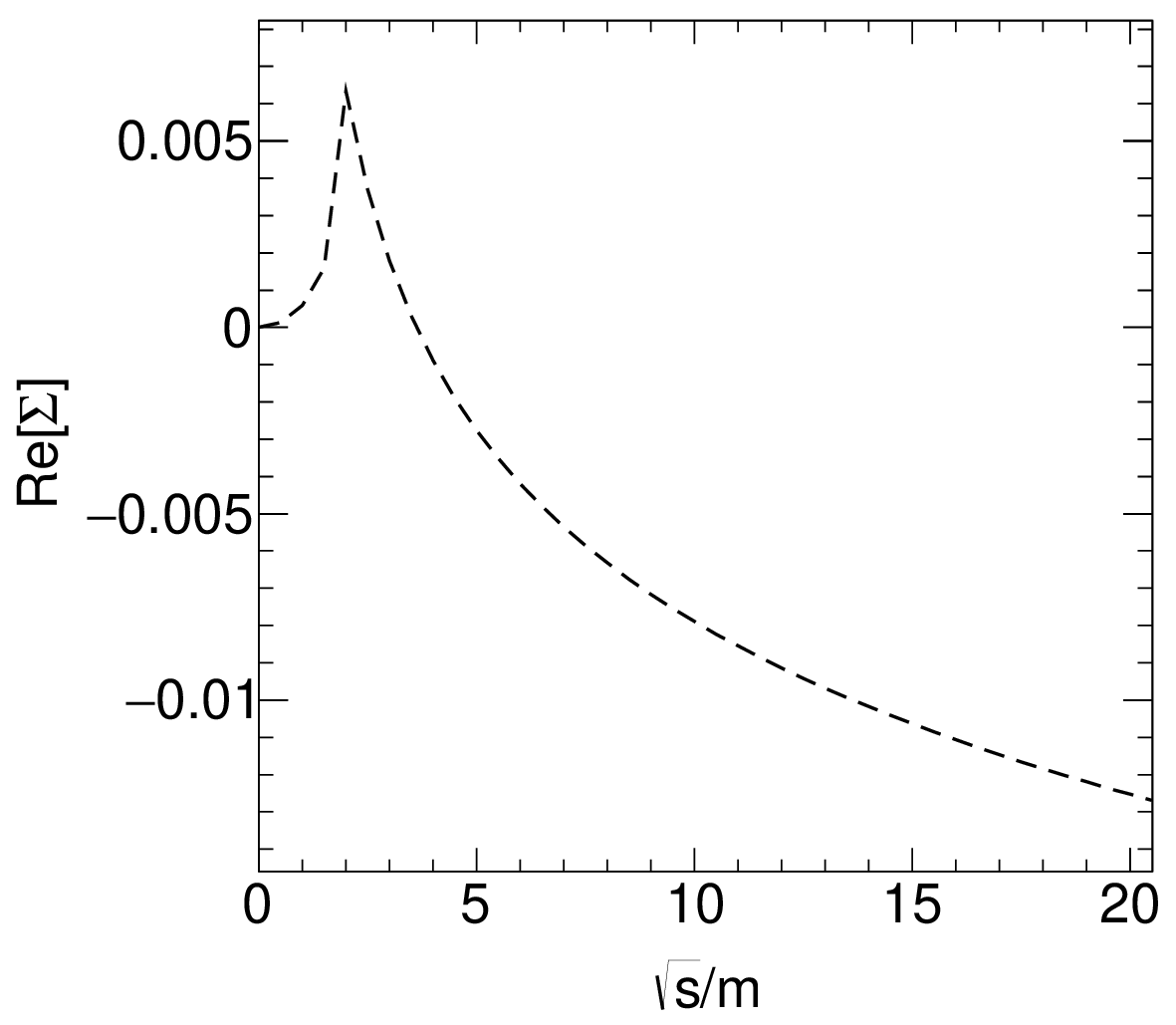}
\includegraphics[width=0.48\textwidth]{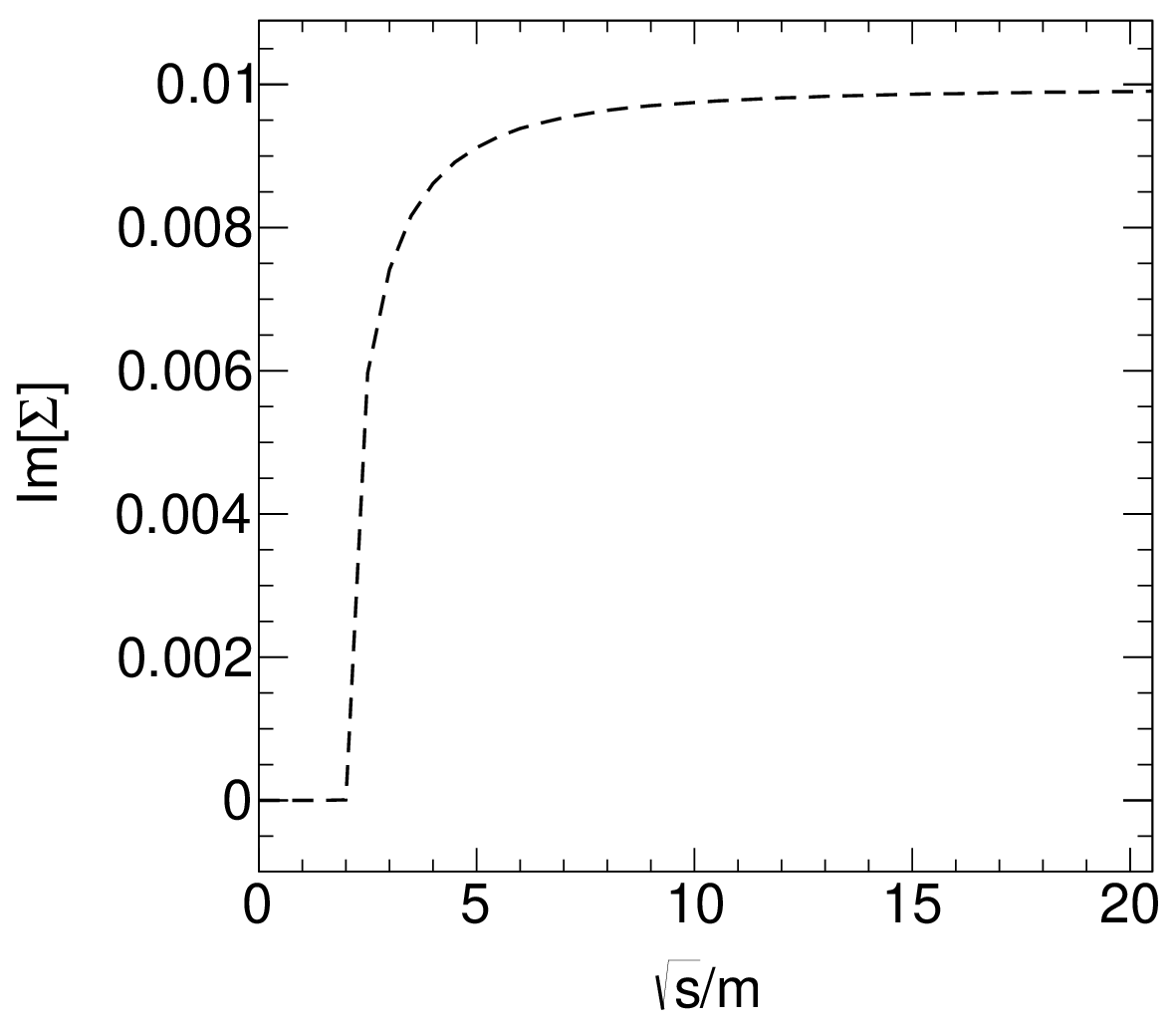}\caption{Real and
imaginary parts of loop function for real $\sqrt{s}/m$.}%
\label{fig:loop_fn}
\end{figure}

The function $\operatorname{Im}\Sigma(s)$ and $\operatorname{Re}\Sigma(s)$
(for real values of $s$) are presented in Fig. \ref{fig:loop_fn}. The real
part rises below threshold, has a cusp at it, then decreases monotonically
and becomes negative for $\sqrt{s}/m$ large enough.
The imaginary part is zero (infinitesimally small) below threshold, then it rises
above it and saturates to the value $1/(32\pi)$ for large $\sqrt{s}/m$. Note, its right-hand-side derivative at
threshold is infinite.

The loop function allows to calculate the unitarized amplitudes in the
$k$-channel as:
\begin{equation}
A_{k}^{U}(s)=\left[  A_{k}^{-1}(s)-\Sigma(s)\right]  ^{-1}.
\end{equation}

All unitarized amplitudes (and consequently phase shifts) with $l=1,2,...$ vanish also at the
unitarized level. The unitarized $s$-wave amplitude and phase shift are
nonzero and take the form:
\begin{equation}
A_{0}^{U}(s)=\left[  A_{0}^{-1}(s)-\Sigma(s)\right]  ^{-1}=\frac{-\lambda
}{1+\lambda\Sigma(s)}\text{ ,}%
\end{equation}%
\begin{equation}
\frac{e^{2i\delta_{0}^{U}(s)}-1}{2i}=\frac{1}{2}\cdot\frac{k}{8\pi\sqrt{s}%
}A_{0}^{U}(s)\text{ .} \label{delta0U}%
\end{equation}
Hence%

\begin{equation}
\delta_{0}^{U}(s)=\frac{1}{2}\arg\left[  1-\frac{1}{8\pi}\sqrt{\frac{m^{2}}%
{s}-\frac{1}{4}}A_{0}^{U}(s)\right]  . \label{eq:ps_unitarized}%
\end{equation}

The scattering length is changed by the unitarization:%
\begin{equation}
a_{0}^{U,\text{SL}}=\frac{1}{2}\frac{A_{0}^{U}(s=4m^{2})}{8\pi\sqrt{4m^{2}}%
}=\frac{1}{2}\frac{1}{16\pi m}\frac{-\lambda}{1+\lambda\Sigma(4m^{2})}%
\end{equation}
Within the used unitarization
\begin{equation}
\Sigma(s=4m^{2})=\frac{1}{16\pi^{2}}\text{ ,}%
\end{equation}
hence it follows that
\begin{equation}
a_{0}^{U,\text{SL}}=\frac{1}{2}\frac{1}{16\pi m}\frac{-\lambda}{1+\frac
{\lambda}{16\pi^{2}}}.
\end{equation}
It is then clear that $a_{0}^{U,\text{SL}}<0$ for $\lambda>0$ (repulsion), and
that $a_{0}^{U,\text{SL}}>0$ for $\lambda\in(\lambda_{c}=-16\pi^{2},0)$
(attraction). However, $a_{0}^{U,\text{SL}}<0$ for $\lambda<\lambda_{c},$
in agreement with the fact that repulsion sets in again. This is due to the
fact that for $\lambda<\lambda_{c}$ a bound-state below threshold emerges, as
we shall show in the next subsection.

Finally one can
calculate $\delta_{0}^{U}(s)$ by using the equivalent expressions%
\begin{align}
\delta_{0}^{U}(s)  &    =\frac
{1}{2}\arcsin\left[  \frac{k}{8\pi\sqrt{s}}\operatorname{Re}\left[  A_{0}%
^{U}(s)\right]  \right] \text{ ,} \label{delta0Uarcsin}\\
\delta_{0}^{U}(s)  &    =\frac{1}{2}\arccos\left[  1-\frac{k}{8\pi\sqrt{s}}\operatorname{Im}%
\left[  A_{0}^{U}(s)\right]  \right]  \text{ .} \label{delta0Uarccos}%
\end{align}
Once the unitarization procedure is employed, the expressions (\ref{delta0Uarcsin}%
), (\ref{delta0Uarccos}), and (\ref{delta0U}) give rise to the same result for
the phase-shift. This is also a useful check of the correctness of our approach.

\subsection{Bound state}
\label{sec:BoundStateFormation}
If $\lambda$ is negative the two scalar particles attract each other. A
natural question is under which condition a bound state emerges. Such a bound
state, denoted as $B$, with mass $M_{B},$ should fulfill the equation (for $s\in (0,4m^{2})$)%
\begin{equation}
A_{0}^{U}(s)^{-1}=\left[  -\lambda^{-1}-\Sigma(s=M_{B}^{2})\right]  =0\text{ .}
\label{bseq}%
\end{equation}
Since $\Sigma(s)$ is real for $s<4m^{2}$ and has a maximum at threshold with $\Sigma(s=4m^{2})=\frac{1}{16\pi^{2}}$ (see
Eq. (\ref{eq:loop})), it turns out
that a bound state is present if
\begin{equation}
\lambda\leq \lambda_{c}=-16\pi^{2}.%
\end{equation}

\begin{figure}[ptb]
\centering
\includegraphics[width=0.48\textwidth]{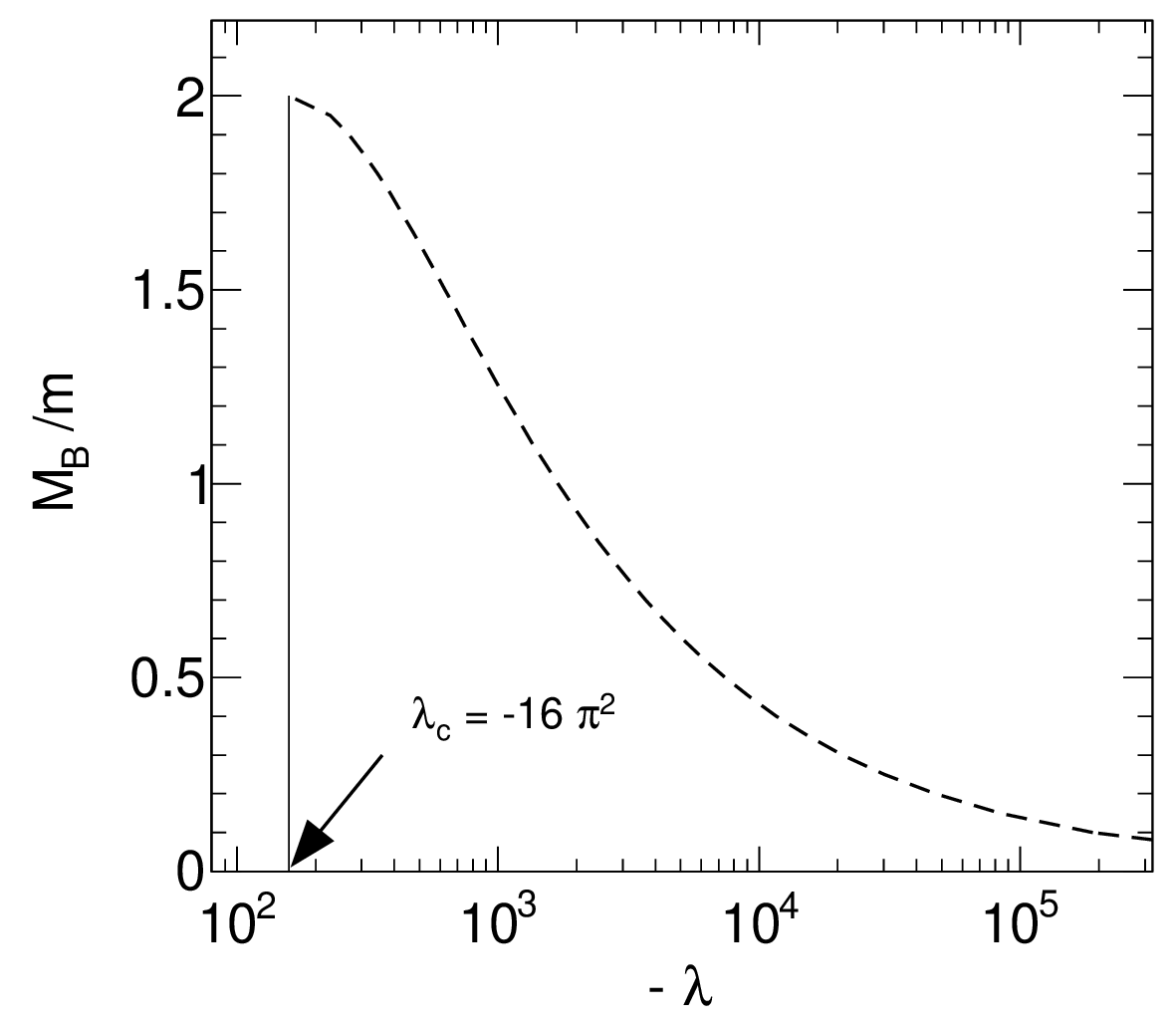}\caption{Mass of the
bound state $M_{B}$ as function of $-\lambda$.}%
\label{fig:mb_vs_lambda}
\end{figure}
The mass $M_{B}$ as a function of $\lambda$, plotted in Fig.
\ref{fig:mb_vs_lambda}, fulfills the conditions:
\begin{align}
M_{B}(\lambda &  =\lambda_{c})=2m\text{ ,}\\
M_{B}(\lambda &  \rightarrow-\infty)=0\text{ .}%
\end{align}
This result also shows the convenience of the employed subtraction scheme: when the
attraction is infinitely strong, the bound state becomes massless. This choice
avoids also the emergence of an additional energy scale into the problem.

Of course, one could perform the study also for different subtraction choices: if
e.g. $\Sigma(0)>0$ the mass $M_{B}$ tends to a finite value for an infinite
negative coupling; if, instead, $\Sigma(0)<0$ a tachyonic mode (instability)
appears for a negative coupling whose modulus is large enough. Alternatively, one could use a
finite cutoff function, but this choice is linked to a non-local Lagrangian
\cite{Burdanov:1996uw,Faessler:2003yf,Giacosa:2004ug}. Yet, all these
possibilities imply that a new energy scale enters into the problem. While
this might be possible, that would introduce an unnecessary complication and
would also spoil the fact that only the mass $m$ entering in\ Eq. (\ref{eq:L})
is the unique energy scale of the system.

In conclusion, the quartic theory of Eq. (\ref{eq:L}) is fully defined only once
its unitarization is settled. The unitarized version of the model together
with the employed subtraction constant
chosen in this work
assures that the model under study is
well defined for any $\lambda$ (positive and negative) and is therefore very
well suited for the study that we aim to do, namely the role of the bound
state in a thermal bath.

\subsection{Behavior of the unitarized phase shift}
\label{sec:UnitarizedPhaseShift}

In order to discuss uniterized phase shift,
an important note on the adopted convention is in
order. We impose that the phase-shift vanishes at threshold:
\begin{equation}
\delta_{0}^{U}(s=4m^{2})=0\text{ },
\end{equation}
regardless on the existence of the bound state below threshold or not. In this
way, the comparison between different curves is better visible. We recall that
often a different convention is used, according to which the phase space at
threshold equals $n_{BS}\pi,$ where $n_{BS}$ is the number of bound states
below threshold \cite{Taylor}. Of course, the choice of the convention
has no impact on the physics. For instance, the Levinson theorem
\cite{Hartle:1965nj,levinson} relates the number of poles below threshold to the difference
of the phase-shift at infinity and at threshold:
\begin{equation}
n_{\text{poles-below-threshold}}=\frac{1}{\pi}\left(  \delta_{0}%
^{U}(s\rightarrow\infty^{2})-\delta_{0}^{U}(s=4m^{2})\right)  \text{ .}%
\end{equation}
This quantity is clearly independent on the choice of an overall constant. In
some cases, the number of poles below threshold equals the number of bound
states, but care is needed, since some unphysical poles may also exist, see
below. 

Similarly, the finite temperature properties studied in the next section
depend on the derivative $d\delta_{0}^{U}(s)/ds,$ which is also independent on
the convention regarding $\delta_{0}^{U}(s=4m^{2}).$We shall also elaborate
more on the behavior of $\delta_{0}^{U}(s)$ in Sec. 3.3.

\begin{figure}[ptb]
\centering
\includegraphics[width=0.48\textwidth]{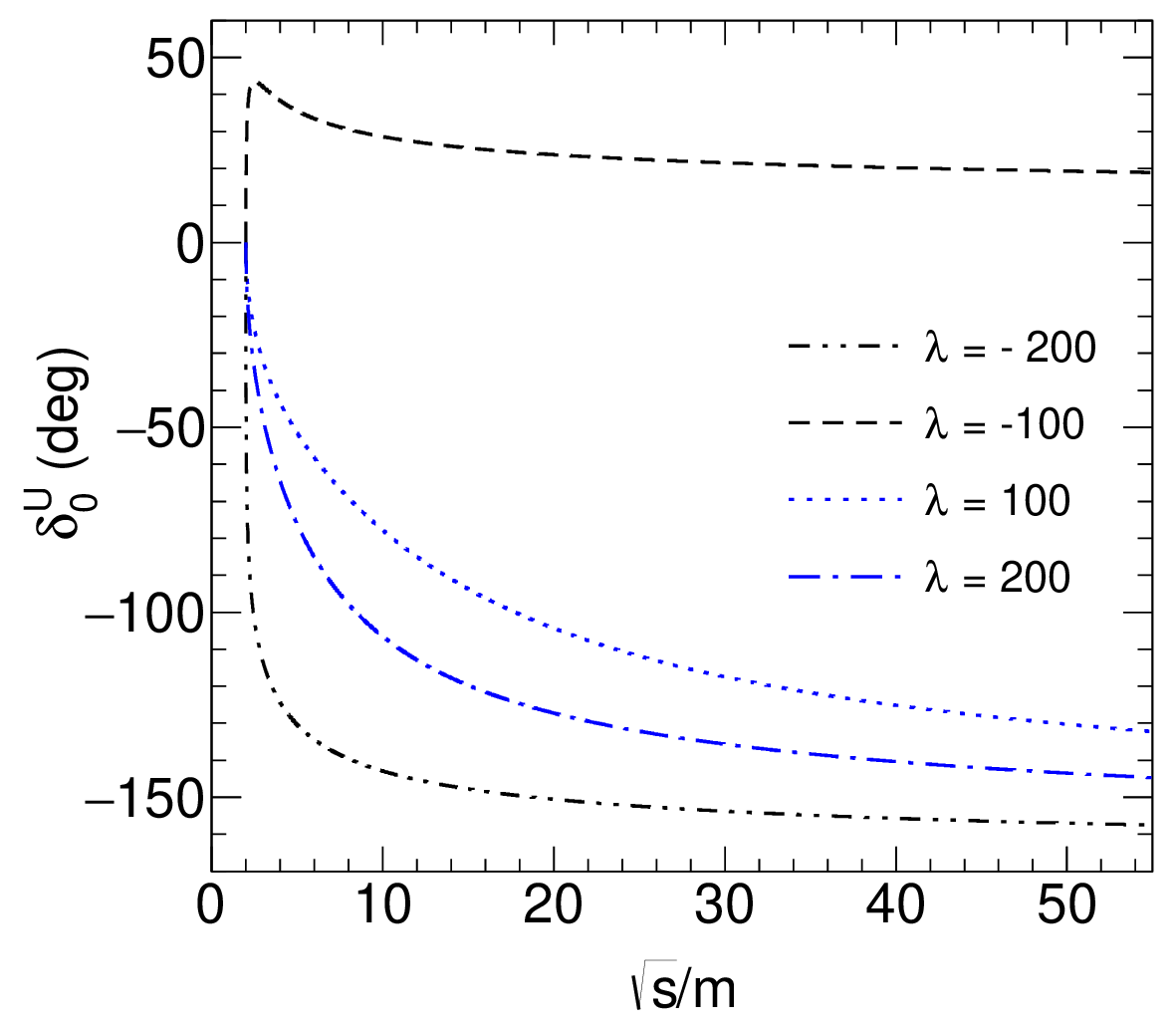}
\caption{Behavior of the unitarized phase-shift $\delta_{0}^{U}$ as function
of $\sqrt{s}/m$ for different values of $\lambda$.}%
\label{fig:ps_loop}%
\end{figure}

Let us now present the behavior of the unitarized phase-shift $\delta_{0}%
^{U}(s)$ in Fig. \ref{fig:ps_loop}. Only for small $\lambda$, the behavior of
$\delta_{0}(s)$ is similar to that of Fig. \ref{fig:phaseshift_treelevel}.
Yet, also in the unitarized case, for $\lambda>0$ the phase-shift and its
derivative are always negative.\ Moreover, the asymptotic value%
\begin{equation}
\delta_{0}^{U}(s\rightarrow\infty)=-\pi\text{ for }\lambda>0 \label{as1}%
\end{equation}
is realized. In addition, the point at which $\delta_{0}^{U}(s=s_{1})=-\pi/2$
is obtained for
\begin{equation}
-\lambda^{-1}-\operatorname{Re}\Sigma(s_{1})=0\text{ ,}%
\end{equation}
where the amplitude becomes purely imaginary with
\begin{equation}
\frac{e^{2i\delta_{0}(s_{1})}-1}{2i}=i.
\end{equation}
The point $s_{1}$ is present for each positive value of $\lambda$ since
$\operatorname{Re}\Sigma(s_{1})$ is unbounded from below. According to the
Levinson theorem \cite{Hartle:1965nj,levinson}, Eq. (\ref{as1}) implies that a
pole below threshold exist. Indeed, for $\lambda>0$ such a pole of the amplitude is present
for a negative value of $s$ that fulfills the very same Eq. (\ref{bseq}), but
of course this pole does not correspond to a physical bound state.

Next, for $\lambda$ negative but belonging to the range $\left(  \lambda
_{c}=-16\pi^{2},0\right)  $, the phase-shift is positive, it rises for small
values of $\sqrt{s}/m,$ it reaches a maximum, and than it bends over
approaching zero for large values of $s$:%
\begin{equation}
\delta_{0}^{U}(s\rightarrow\infty)=0\text{ for }\lambda\in\left(  \lambda
_{c},0\right)  .
\end{equation}
This is also in agreement with the Levinson's theorem, since no pole below
threshold appears.

Finally, for $\lambda<\lambda_{c}$ the phase $\delta_{0}^{U}(s)$ is negative
and approaches $-\pi$:
\begin{equation}
\delta_{0}^{U}(s\rightarrow\infty)=-\pi\text{ for }\lambda<\lambda_{c}\text{,}%
\end{equation}
in accordance with Levinson's theorem, since a pole for $s=M_{B}^{2}$ exists.
Also in this case, there is a certain value $s=s_{1}$ at which the phase is
$-\pi/2.$

\begin{figure}[ptb]
\centering
\includegraphics[width=0.48\textwidth]{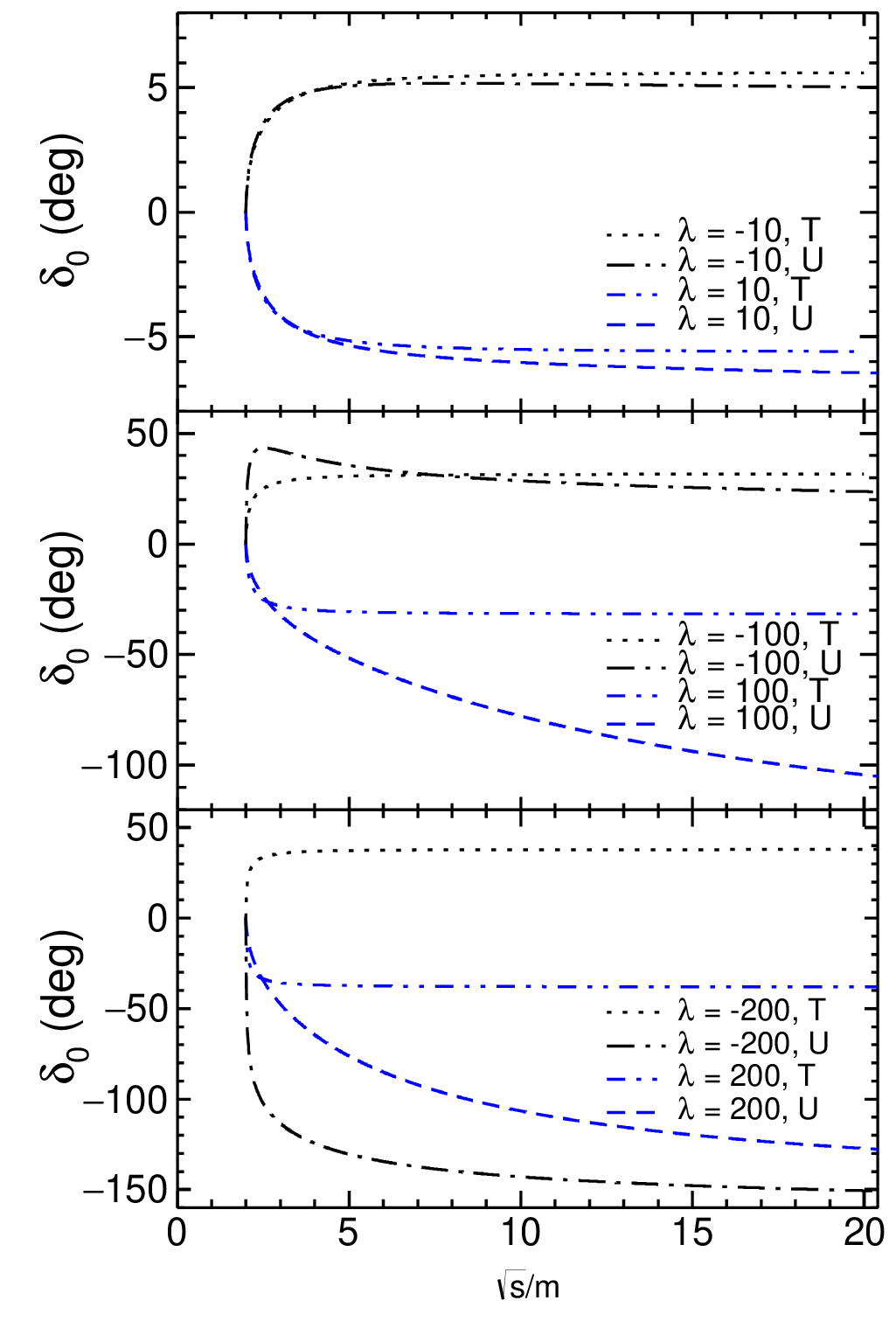}
\caption{Comparison of tree level (T) and unitarized (U) phase shifts as function
of $\sqrt{s}/m$ for different values of $\lambda$.
As we discuss in the text, the phase-shift is chosen to
vanish at threshold ($\delta_{0}^{U}(s=4m^{2})=0$ ), independently on the
value of $\lambda.$ In this way it is easy to compare the behavior of the
phase-shift for different values of $\lambda$, even when a bound state
emerges. This choice does not affect the physics. 
}%
\label{fig:ps_comparison}%
\end{figure}

In Fig. \ref{fig:ps_comparison} we compare the tree level (T) and the unitarized (U) phase shifts.
The top panel shows the results when $\lambda$ is small ($\pm 10$). 
The qualitative behavior of the phase shifts for both cases is similar for all $\sqrt{s}/m$
shown in the figure. When $\sqrt{s}/m$ is small ($<4$), the tree-level and unitarized results are very close to each other, then a discrepancy is appreciable
at larger values of $\sqrt{s}/m$. 
In the middle panel  we show a similar
comparison for $\lambda = \pm 100$.  In this case the unitarized phase shifts differ significantly from the tree-level ones. 
For  $\lambda = -100$,  the unitarized phase 
shift  first increases sharply
for increasing $\sqrt{s}/m$, reaches a maximum, and then starts decreasing.
The magnitude of the unitarized phase shift is larger than that at tree-level
at low $\sqrt{s}/m$, but becomes smaller at large $\sqrt{s}/m$. 
For $\lambda = +100$ both the tree level and the unitarized phase shift decrease for increasing $\sqrt{s}/m$. However, the decrease is much steeper for the unitarized phase shift.
The bottom panel of Fig. \ref{fig:ps_comparison} shows the choice
$\lambda = \pm 200$. For $\lambda = 200$ the comparison of the tree-level and unitarized phase shift is similar to that of $\lambda = 100$. 
However, the behavior of unitarized phase shift  for  $\lambda = -200$ is completely different from 
the tree level one. While tree-level phase shift is positive, the unitarized phase-shift is negative because
 $\lambda<\lambda_c$. Correspondingly, in this case the bound state that dominates the near-threshold phenomenology is built.

\section{Thermodynamical properties of the theory}

\label{sec:finiteTemp}

We now consider the thermodynamics (TD) of the system at nonzero temperature.
We first discuss the pressure of the system by using the phase-shift approach
at tree-level, in which no bound state is present, and then at the unitarized
one-loop level. Within the latter scheme, we study the contribution to the TD
of an emerging bound state when the attraction is large enough to form it ($\lambda\leq\lambda_c$).

\subsection{Pressure without the bound state: tree-level results}
\label{sec:P_treelevel}

The non-interacting part of the pressure for a gas of particles with mass $m$ reads:%

\begin{equation}
P_{\varphi\text{,free}}=-T\int_{k}\ln\left[  1-e^{-\beta\sqrt{k^{2}+m^{2}}%
}\right]  \text{ .}%
\end{equation}

In the $S$-matrix formalism
\cite{Dashen:1969ep,Venugopalan:1992hy,Broniowski:2015oha,Lo:2017ldt,Lo:2017sde,Lo:2017lym,Dash:2018can,Dash:2018mep,Lo:2019who}%
, the interacting part of the pressure is related to the derivative of the
phase shift with respect to the energy by the following relation:%

\begin{equation}
P_{\varphi\varphi\text{-int}}=-T\int_{2m}^{\infty}dx\frac{2l+1}{\pi}\sum
_{l=0}^{\infty}\frac{d\delta_{l}(s=x^{2})}{dx}\int_{k}\ln\left[
1-e^{-\beta\sqrt{k^{2}+x^{2}}}\right]  \text{ ,}%
\end{equation}
where $x=\sqrt{s}$. In our specific case, only the $s$-wave contribution is nonzero:
\begin{equation}
P_{\varphi\varphi\text{-int}}=-T\int_{2m}^{\infty}dx\frac{1}{\pi}\frac
{d\delta_{0}(s=x^{2})}{dx}\int_{k}\ln\left[  1-e^{-\beta\sqrt{k^{2}+x^{2}}%
}\right]  \text{ .}%
\end{equation}
Then, the total tree-level pressure (obviously in the absence of a bound state,
since at tree-level it cannot be generated) is given by
\begin{equation}
P_{tot}=P_{\varphi\text{,free}}+P_{\varphi\varphi\text{-int}}\text{ (at
tree-level). }%
\end{equation}

The previous equations show that we can evaluate the pressure at $T>0$ by using solely the phase shift evaluated in the vacuum.
Of course, all other relevant thermodynamic quantities of the thermal system  (such as energy and entropy densities, etc.)
can be determined once the pressure is known.

\begin{figure}[ptb]
\centering
\includegraphics[width=0.48\textwidth]{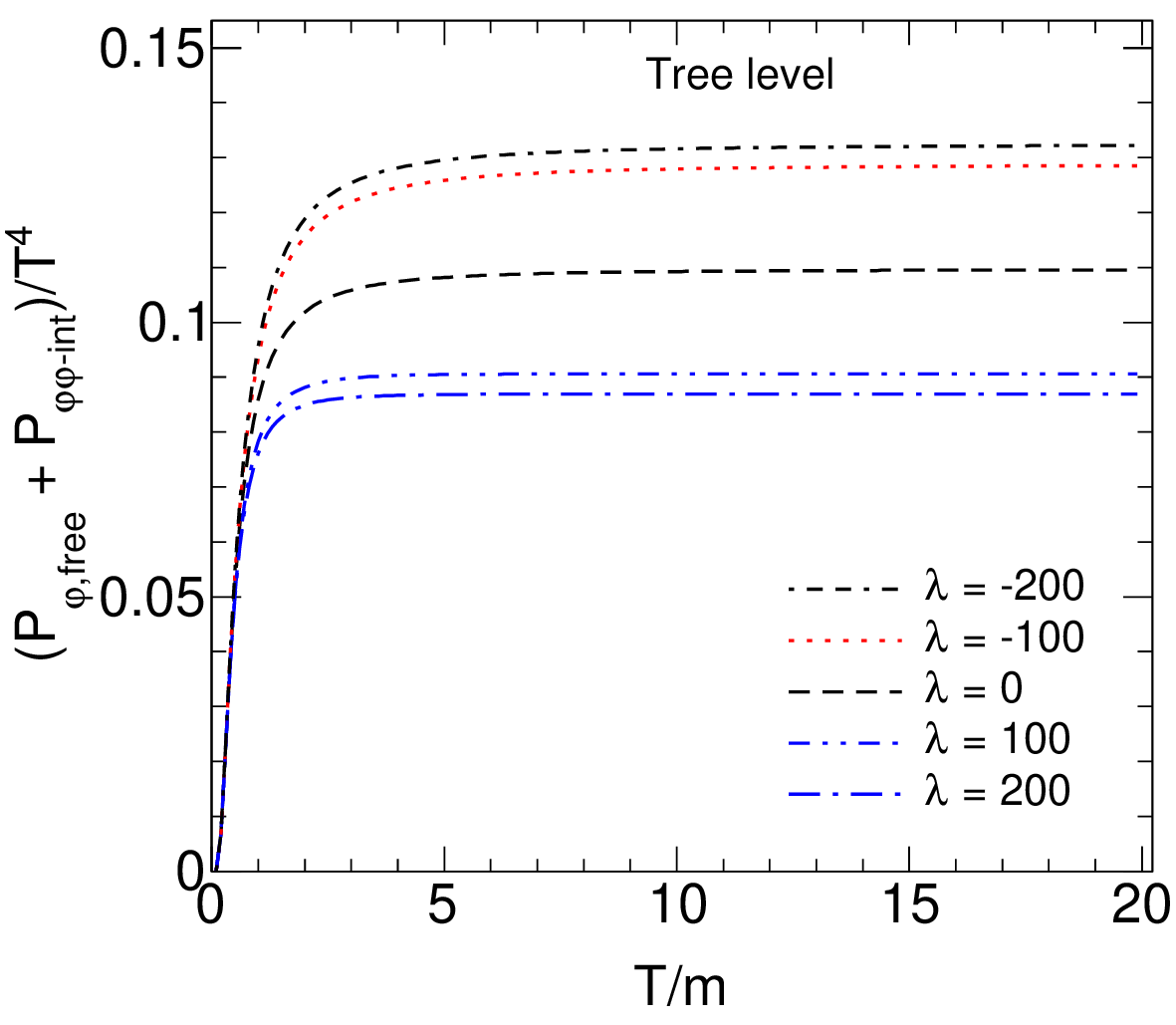}\caption{Tree-level plots of the normalized pressure as function of $T/m$ for different values of $\lambda$.}%
\label{fig:P_T_treelevel}
\end{figure}

The temperature dependence of the corresponding pressure $(P_{\varphi
,free}+P_{\varphi\varphi-int})/T^{4}$ is shown in Fig. \ref{fig:P_T_treelevel}%
. The $\lambda=0$ line corresponds to the pressure of a free gas $P_{\varphi,free}/T^{4}$ that for large $T/m$ saturates towards the massless limit
($P_{\varphi,free}/T_{m=0}^{4}=\pi^{2}/90$). For positive (negative) $\lambda
$, the tree-level repulsive (attractive interaction implies that the pressure is smaller (larger) than the non-interacting case, but never exceed 0.5. As we shall see, the unitarization enhances the contribution of the interaction. 

\begin{figure}[ptb]
\centering
\includegraphics[width=0.48\textwidth]{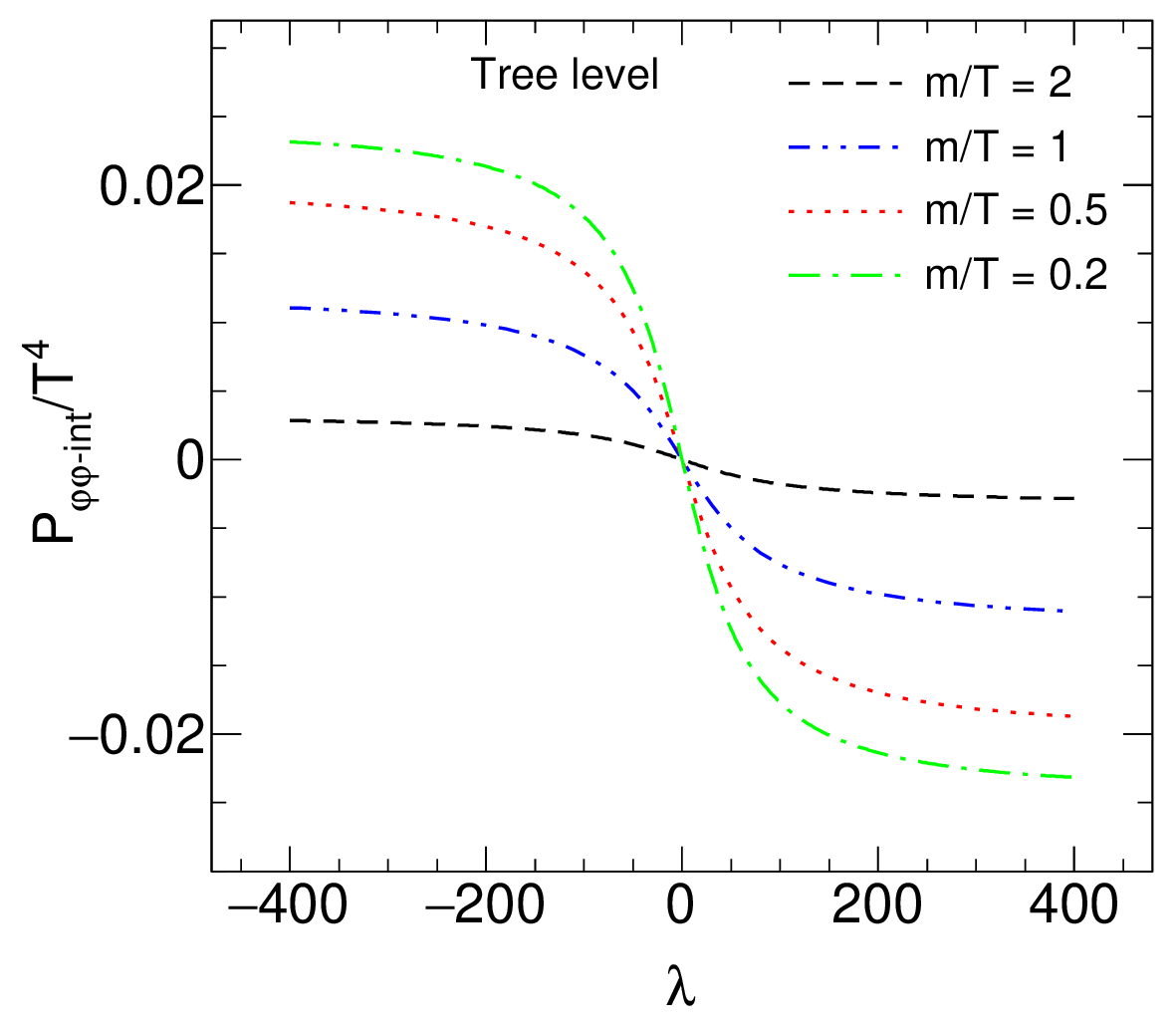}
\includegraphics[width=0.48\textwidth]{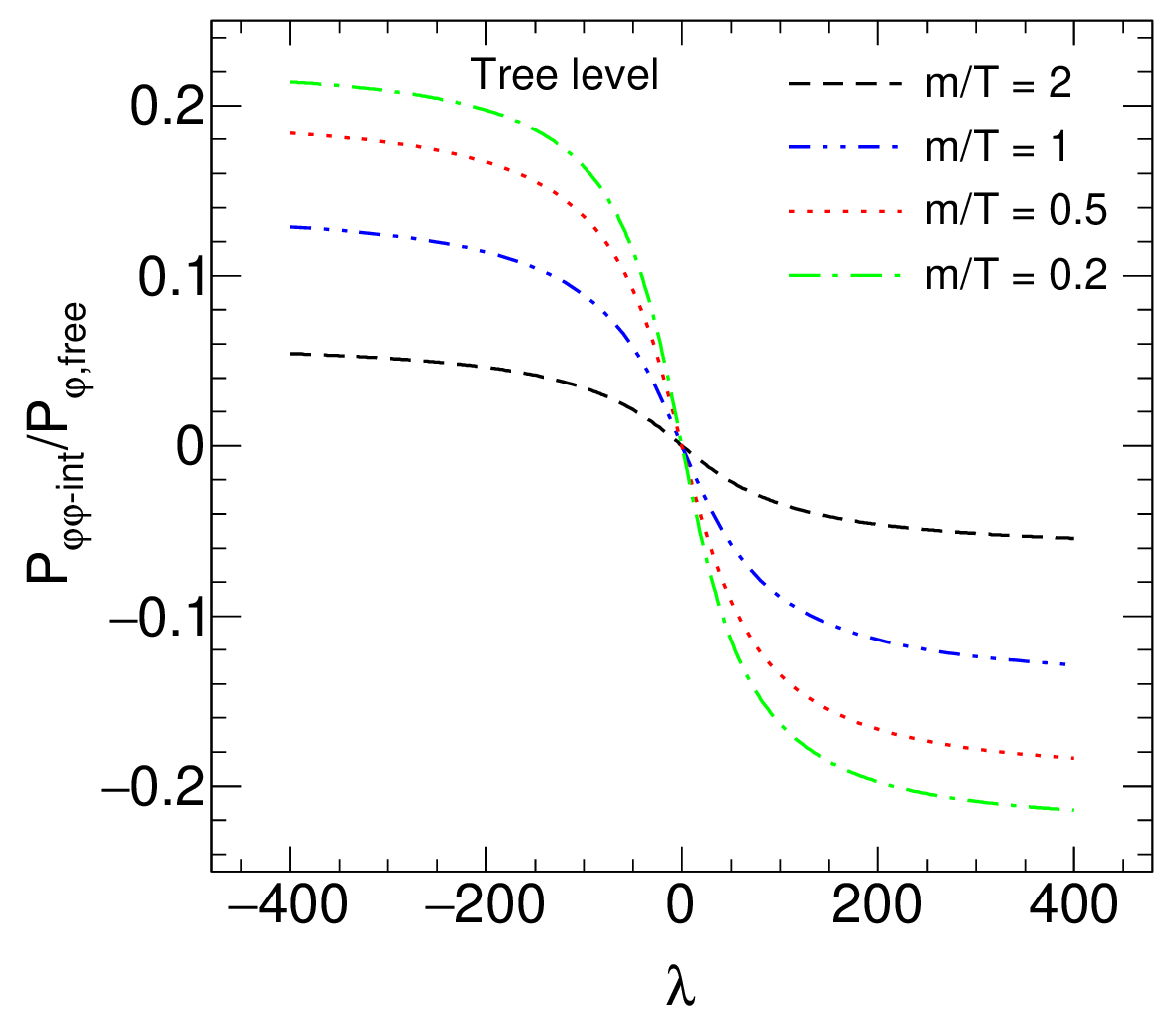}\caption{The
left panel shows the normalized pressure at tree-level as function of
$\lambda$ for different temperatures. The right panel shows the ratio of the interacting pressure over the pressure of a free gas as function of $\lambda$ for different temperatures.}%
\label{fig:P_vs_lambda_treelevel}
\end{figure}

Next, in Fig. \ref{fig:P_vs_lambda_treelevel} we study $P_{\varphi\varphi
-int}/T^{4}$ and $P_{\varphi\varphi-int}/P_{\varphi,\text{free}}$ as function
of $\lambda$ for four different $m/T$ ratios $2,$ $1,$ $0.5$ and $0.2$. One
can see that near $\lambda=0$, $P_{\varphi\varphi
-int}/T^{4}$ changes rapidly with $\lambda$, but then saturates at large
values of $\lambda$. In the right panel, one can see that all the curves of
the function $P_{\varphi\varphi-int}/P_{\varphi,\text{free}}$ cross the origin
at $\lambda=0,$ which is expected since there is no interaction at
$\lambda=0$. Further, it can be seen that the effect of the interaction is larger
both for large $\lambda$ and/or low $m/T$.

\subsection{Pressure without the bound state: unitarized results}
\label{sec:P_wo_BS}
When including the unitarization procedure explained in Sec.
\ref{sec:Unitarization}, the interaction contribution to the pressure is
obtained by using the unitarized phase-shift into the $S$-matrix formalism:%

\begin{equation}\label{eq:Pint_swave}
P_{\varphi\varphi\text{-int}}^{U}=-T\int_{2m}^{\infty}dx\frac{1}{\pi}%
\frac{d\delta_{0}^{U}(s=x^{2})}{dx}\int_{k}\ln\left[  1-e^{-\beta\sqrt
{k^{2}+x^{2}}}\right]  \text{ .}%
\end{equation}
Then, the total pressure (in the absence of a bound state) is given by
\begin{equation}
P_{tot}^{U}=P_{\varphi\text{,free}}+P_{\varphi\varphi\text{-int}}^{U}\text{
(unitarized, for }\lambda>\lambda_{c}).
\end{equation}

\begin{figure}[ptb]
\centering
\includegraphics[width=0.48\textwidth]{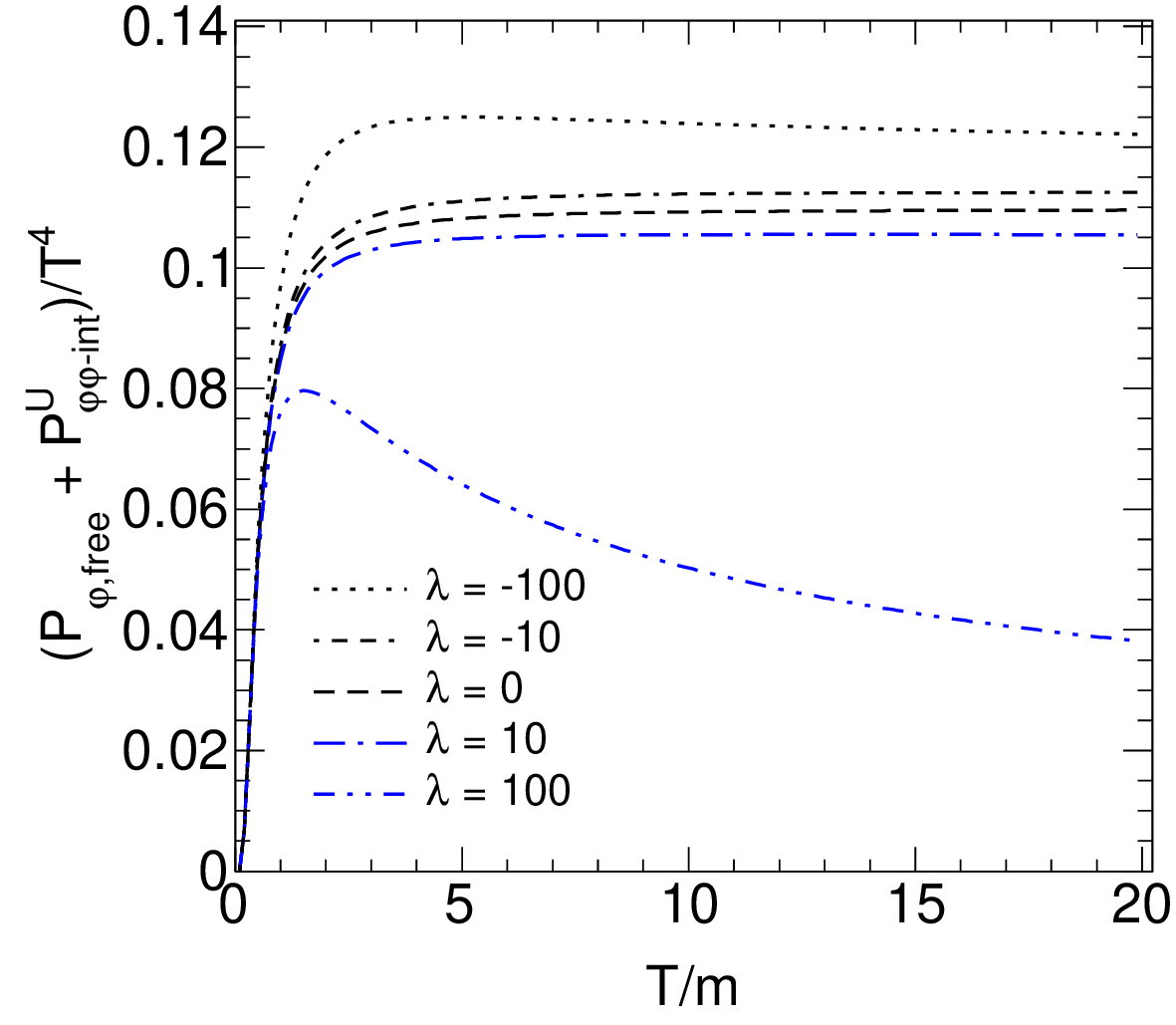}
\caption{Temperature dependence of the normalized pressure in the unitarized case for $\lambda=\pm10$ and for $\lambda=\pm100$.}%
\label{fig:P_vs_T_uniterized}
\end{figure}

Figure~\ref{fig:P_vs_T_uniterized} shows the temperature dependence of pressure in the unitarized case. [Note, no bound state contribution is present here since all the considered values of the coupling  $\lambda$ are larger than $\lambda_c$.] 
For small $\lambda$ ($\pm 10$), the normalized pressure
saturates at large $T/m$. 

Yet, for $\lambda=\pm100$ the normalized pressure as function of the temperature is quite different from the non-interacting case, since it reaches a maximum for a finite value of the temperature.  
In general, this figure shows that for large values of $\lambda$ and for large temperatures, the unitarized result is sizably different from the tree level result reported in Fig. \ref{fig:P_T_treelevel}.

\subsection{The general case: Inclusion of the bound state, formal aspects,
and numerical results}
\label{sec:P_w_BS}

The crucial question of the present work is how to include the effect of the emergent bound state $B$ in the
thermodynamics. The easiest way is to add to the pressure of the system the pressure of of mass $M_{B}$ as:%

\begin{equation}
P_{B}\overset{\text{ }}{=}-\theta(\lambda_{c}-\lambda)T\int_{k}\ln\left[
1-e^{-\beta\sqrt{k^{2}+M_{B}^{2}}}\right]  \label{pb},%
\end{equation}
where the theta function takes into account that for $\lambda>\lambda_{c}$
there is no bound state $B.$ Of course, $M_{B}$ is itself also a function of
$\lambda,$ see Eq. (\ref{bseq}) and Fig. \ref{fig:mb_vs_lambda}.

Within this context, the full (unitarized) pressure looks like
\begin{equation}
P_{tot}^{U}=P_{B}+P_{\varphi\text{,free}}+P_{\varphi\varphi\text{-int}}%
^{U}\text{ (unitarized, for any }\lambda\text{).}%
\end{equation}
Quite remarkably, $P_{tot}^{U}$ turns out to be a continuous function of
$\lambda$, even if $P_{B}$ is not continuous at $\lambda=\lambda_{c}$ since it jumps
abruptly from $0$ to a certain finite value. Yet, the quantity $P_{\varphi
\varphi\text{-int}}^{U}$ is also not continuous in such a way to compensate
the previous jump, see below.

The issue is if the inclusion of $P_{B}$ as in Eq. (\ref{pb}) is correct.
To study this point, we discuss how the contribution of the bound state can be
formally included into the phase-shift analysis, showing that the simple
prescription of adding one additional state to the thermodynamics is correct
and the result is independent on the residuum of the pole of the bound state.

In order to show these features, let us first modify Eq. (\ref{delta0U}) by
extending its validity also below the threshold. To this end we consider
\begin{equation}
\frac{e^{2i\delta_{0}^{U}(s)}-1}{2i}=\operatorname{Im}\Sigma(s)\cdot A_{0}%
^{U}(s)\text{ ,}%
\end{equation}
where $\operatorname{Im}\Sigma(s)$ is given by\ Eq. (\ref{imsigma}). Clearly,
above threshold nothing changes. On the other hand, below threshold we get the
following expression:
\begin{equation}
\frac{e^{2i\delta_{0}^{U}(s)}-1}{2i}=\varepsilon A_{0}^{U}(s)=\frac
{\varepsilon}{A_{0}^{-1}(s)-\Sigma(s)}%
\end{equation}
Note, if $\varepsilon$ is set strictly to zero, we get obviously zero. If
there is no pole below threshold, $\delta_{0}^{U}$ is an infinitesimally small
number, that can be set to zero and has no effect in the description of the system.

Next, let us assume that a bound state below threshold appears: $A_{0}%
^{-1}(s)-\Sigma(s)=0$ for $s=M_{B}^{2}\in(0,4m^{2}).$ In this case, we have
(below threshold):
\begin{equation}
\frac{e^{2i\delta_{0}^{U}(s)}-1}{2i}=\frac{\varepsilon}{-Z^{-1}(s-M_{B}%
^{2})+i\varepsilon}\text{ (for }0<s<4m^{2}\text{)}%
\end{equation}
where
\begin{equation}
Z=\frac{1}{\Sigma^{\prime}(s=M_{B}^{2})}\text{ .}%
\end{equation}
Using the expression for the phase-shift of Eq. (\ref{delta0Uarccos}) we
find:
\begin{equation}
\delta_{0}^{U}(s)=\frac{1}{2}\arccos\left[  1-\frac{2\varepsilon^{2}}{\left[
Z^{-1}\left(  s-M_{B}^{2}\right)  \right]  ^{2}+\varepsilon^{2}}\right]
\text{ (for }0<s<4m^{2}\text{).}%
\end{equation}

For $0<s<M_{B}^{2}$ the argument of the $\arccos$ is $1$ (for an arbitrary
small $\varepsilon$), then unitarized phase shift $\delta_{0}^{\text{U}}=n\pi,$ where $n$ is
an integer. We recall that it in this work we require that $\delta_{0}^{U}(s)$
vanishes at threshold:%
\begin{equation}
\delta_{0}^{U}(s=4m^{2})=0\text{ }.
\end{equation}
By assuming that there is a single pole below threshold, for $s<M_{B}^2$ it is
useful to impose that $n=-1$:
\begin{equation}
\delta_{0}^{U}(0<s<M_{B})=-\pi\text{ for }0<s<M_{B}^{2}\text{ .}%
\end{equation}
Next, we notice that for $s=M_{B}^{2}$, the argument equals to $1-\frac
{2\varepsilon^{2}}{\varepsilon^{2}}=-1,$ therefore $\delta_{0}^{U}=\frac{n}{2}\pi$ for
this particular choice of $s.$

The function $\delta_{0}^{U}(s=x^{2})$ must be (for a finite $\varepsilon,$
even if arbitrarily small) a continuous and differentiable function. Hence, it
follows that
\begin{equation}
\delta_{0}^{U}(s=M_{B}^{2})=-\frac{\pi}{2}\text{ }.
\end{equation}
Moreover, for any value of $M_{B}^{2}<s<4m^{2}$ we have
\begin{equation}
\delta_{0}^{U}(M_{B}^{2}<s<4m^{2})=0\text{ }.
\end{equation}
We may then conclude that for $s\in(0,4m^{2})$, alias for $x\in(0,2m)$, the
phase shift takes the form:%
\begin{equation}
\delta_{0}^{U}(x=\sqrt{s})=-\pi+\pi\theta(x-M_{B})\text{ .}%
\end{equation}
In this way we obtain the desired result:%
\begin{equation}
\frac{1}{\pi}\frac{d\delta_{0}^{U}(x)}{dx}=\delta(x-M_{B})\text{ .}%
\end{equation}

Quite interestingly, this result is independent on the residue of the pole $Z.$ The bound state counts
always as $1,$ showing that the corresponding density of states is given by
\begin{equation}
n_{B}\overset{\text{ }}{=}\theta(\lambda_{c}-\lambda)\int_{k}\left[
e^{-\beta\sqrt{k^{2}+M_{B}^{2}}}-1\right]  ^{-1}\text{ ,}%
\end{equation}
in agreement with thermal models.

\begin{figure}[h]
\centering
\includegraphics[width=0.48\textwidth]{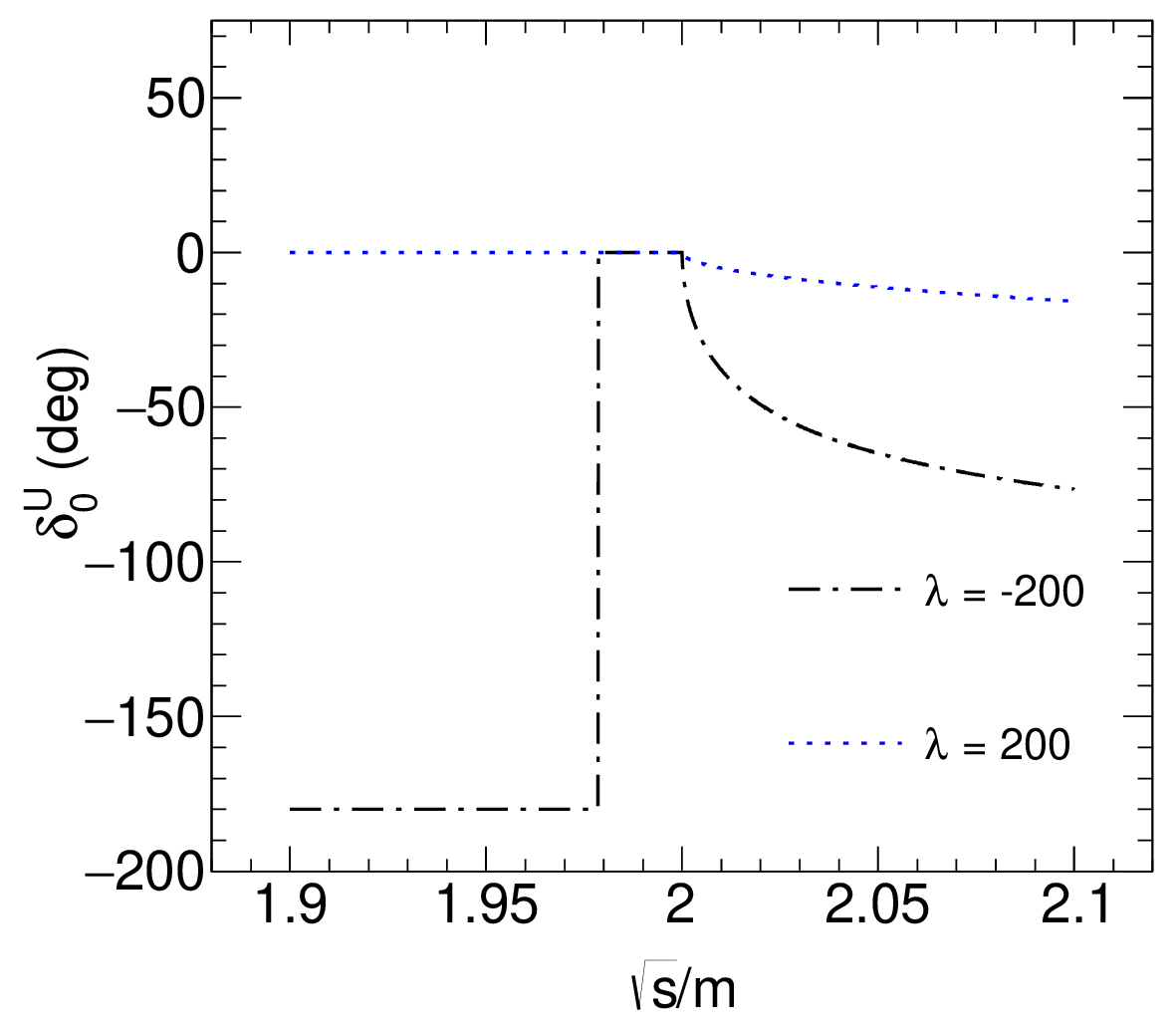}
\includegraphics[width=0.48\textwidth]{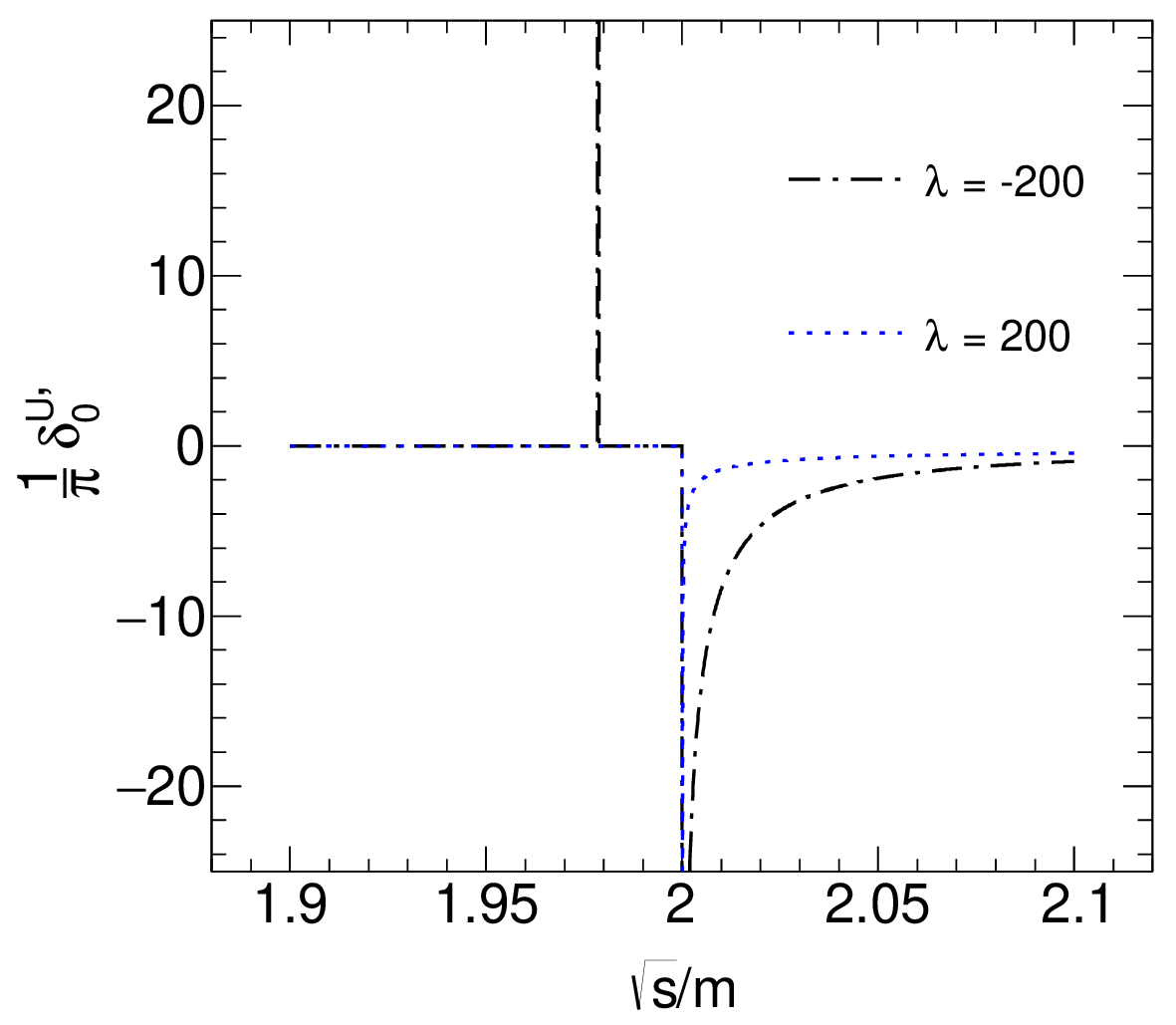}
\caption{Left panel shows the energy dependence of the unitarized phase shift for $\lambda=\pm200$. Right panel shows the derivative of the corresponding phase shifts.}%
\label{fig:ps_bs}
\end{figure}

 In order to understand better the behavior of the
phase-shift, we show in the left panel of Fig. \ref{fig:ps_bs} the behavior of the
unitarized phase shift below threshold  for two different $\lambda$ values,
one below and another above the critical value.  For $\lambda=200>\lambda_c$, the phase
shift is simply zero below the threshold and decreases with the increase of
$\sqrt{s}/m$, whereas, for $\lambda=-200<\lambda_c$, the phase shift is $-\pi$
(according to our convention) below the mass of the bound state ($M_B/m\sim1.98$). The phase shift jumps to zero for $\sqrt{s}=M_{B}$ and
remains zero upto the threshold. This jump of phase shift is due to the
formation of the bound state. Above threshold the phase shift decreases with
the increase of $\sqrt{s}/m$. 

The right panel of Fig. \ref{fig:ps_bs} shows the energy dependence of the derivative of the
phase shift. For $\lambda=200$, the derivative of the phase shift is zero
below threshold. Above threshold this quantity is negative and its
magnitude increases with the increase of $\sqrt{s}/m$. For $\lambda=-200$,
there is a delta function at $\sqrt{s}=M_{B}$, which is responsible for the
inclusion of the bound state in the phase-shift formalism. Indeed, as shown in
Eq. \ref{eq:Pint_swave}, the pressure depends on the derivative
of the phase shift, hence the functions depicted in the right panel of Fig. \ref{fig:ps_bs}
represent the two-particle energy weight.

One can also understand form the plots in Fig. \ref{fig:ps_bs} that, using the more common
convention according to which the phase-shift equals $\pi$ at threshold when a
bound state is present, would amount to consider $\delta_{0}^{U}(s)+\pi$ for
$\lambda=-200$ in the left panel, while the right panel would remain
unchanged. This result shows that the choice of the phase-shift value at
threshold does not affect the thermodynamics (as well as any other physical
property), as it should.

Finally, we turn to the thermodynamics of the system. The pressure contributions from the bound state and from the
interaction can be described by the following expression%
\begin{equation}
P_{\varphi\varphi\text{-int-tot}}=P_{\varphi\varphi\text{-int}}^{U}%
+P_{B}=-T\int_{0}^{\infty}dx\frac{1}{\pi}\frac{d\delta_{0}^{U}(s=x^{2})}%
{dx}\int_{k}\ln\left[  1-e^{-\beta\sqrt{k^{2}+x^{2}}}\right]  \text{ ,}%
\end{equation}
where the lower bound of the integral is now set to zero. If the bound state
is present, it is \textit{automatically} taken into account (independently on
the binding energy).

Next, we discuss the numerical result in presence of a bound state. As we have
already mentioned, the formation of bound state is possible when $\lambda$
is less than the critical value $\lambda_{c}=-16\pi^{2}$.

\begin{figure}[ptb]
\centering
\includegraphics[width=0.48\textwidth]{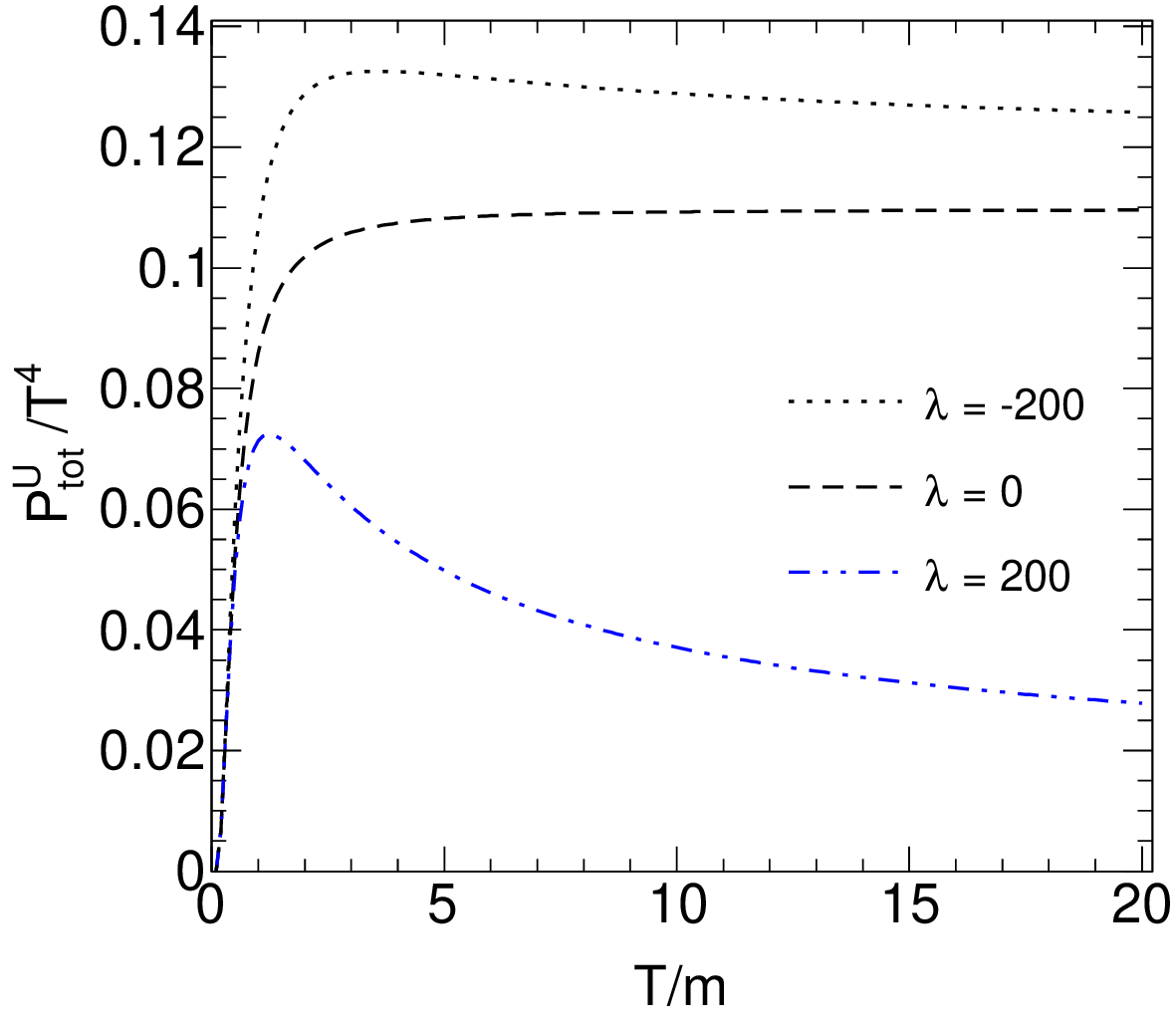}
\caption{Temperature dependence of the normalized pressure in the unitarized case for $\lambda=\pm200$.}%
\label{fig:P_tot_vs_T_uniterized}
\end{figure}

Figure \ref{fig:P_tot_vs_T_uniterized} shows the temperature dependence 
of the normalized total pressure for $\lambda = \pm200$. For the value $\lambda = -200$ (which is less than $\lambda_c$) the bound state is present and, as expected, the total normalized pressure is larger than that of non-interacting particle. For the value $\lambda = 200$ the total pressure is strongly reduced. Yet, in general, the
 qualitative behavior of the curves for $\lambda = \pm 200$ is quite similar to those for $\lambda = \pm 100$ depicted in Fig. \ref{fig:P_vs_T_uniterized}.

\begin{figure}[ptb]
\centering
\includegraphics[width=0.48\textwidth]{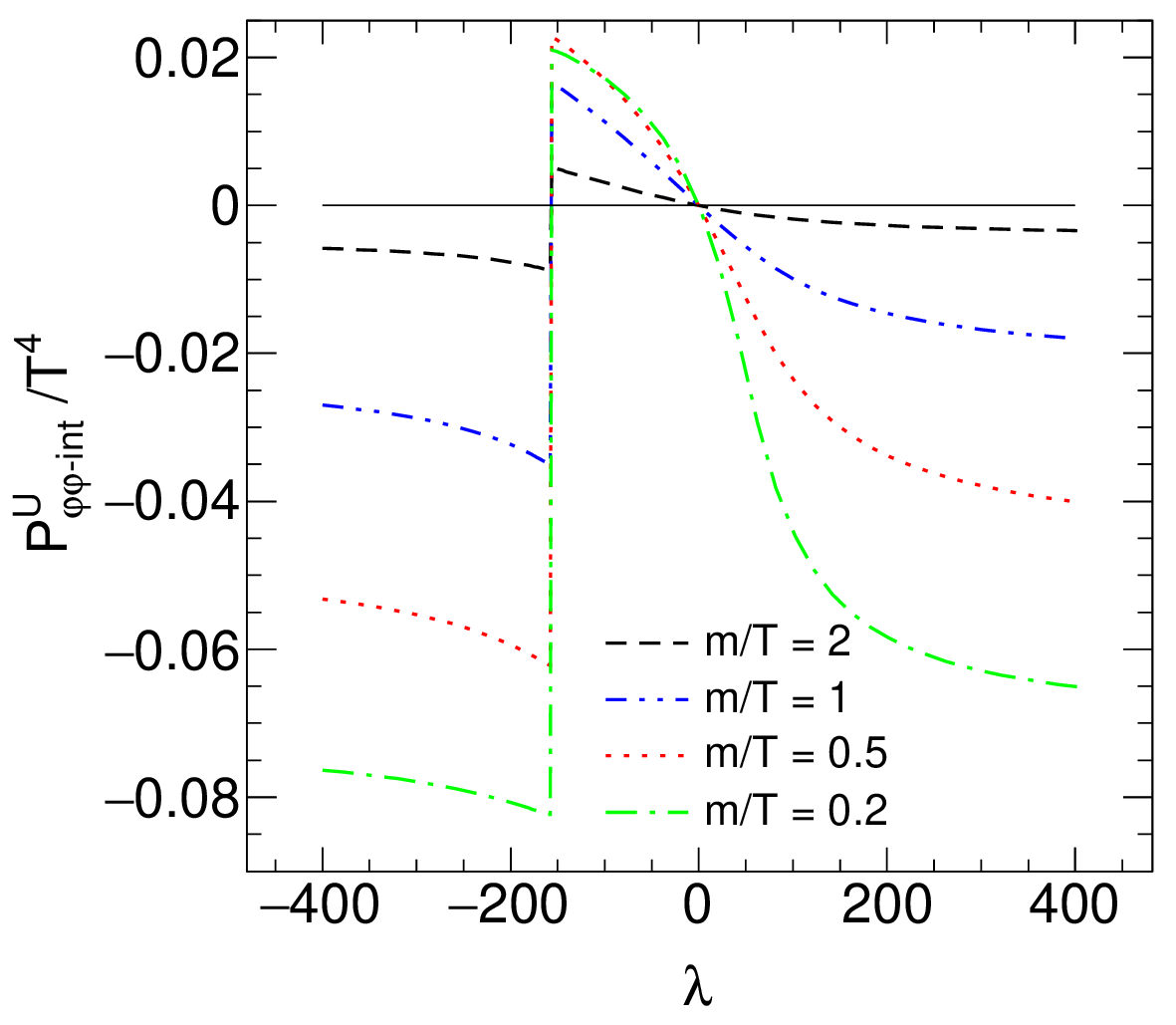}
\includegraphics[width=0.48\textwidth]{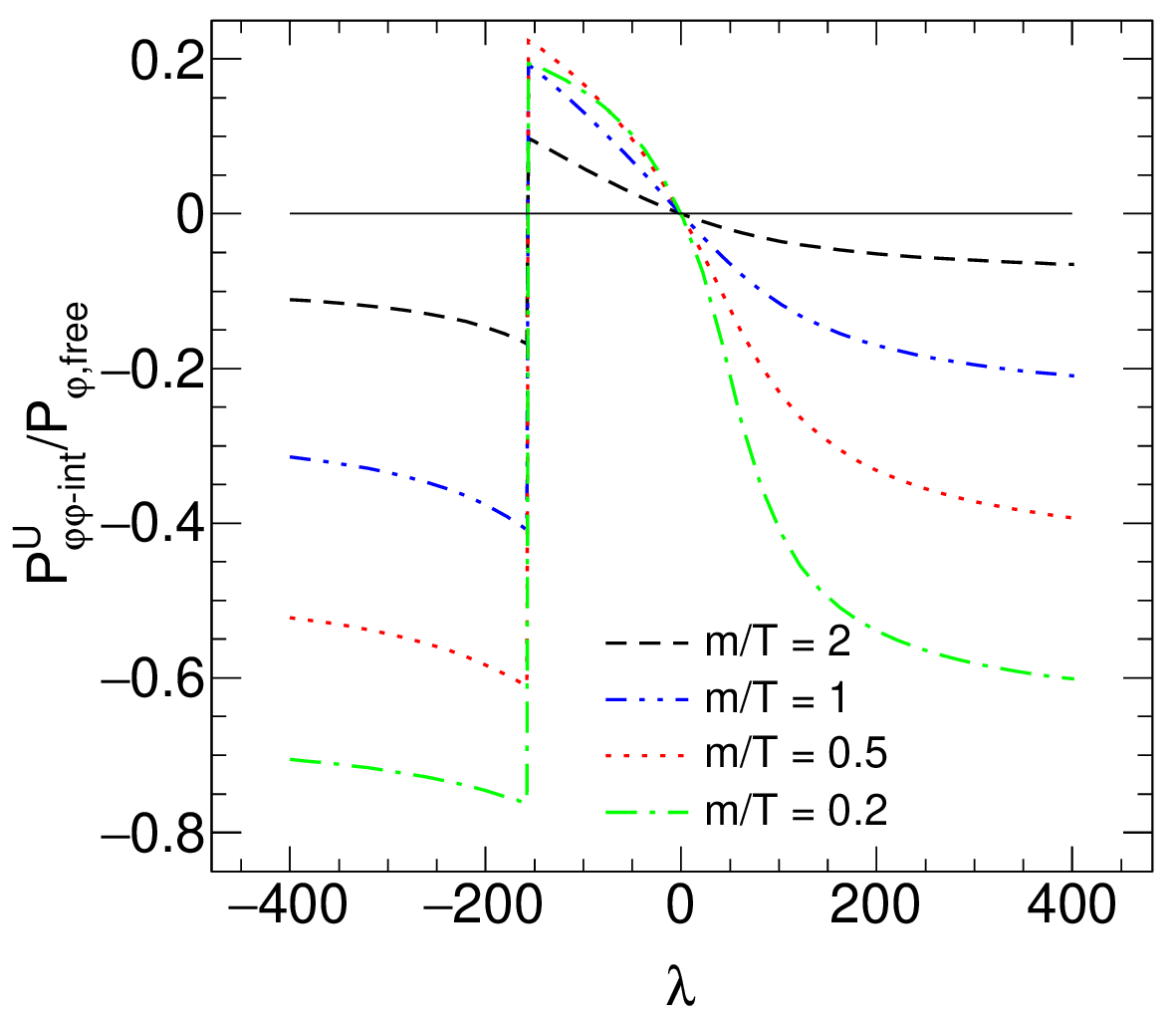}\caption{The
left panel shows the behavior of the interacting part of the normalized pressure with $\lambda$.
The right panel shows the $\lambda$-dependence of the interacting part of the pressure relative to that of a free gas of particles with mass $m$.}%
\label{fig:p_lambda_uni}
\end{figure}

\begin{figure}[ptb]
\centering
\includegraphics[width=0.48\textwidth]{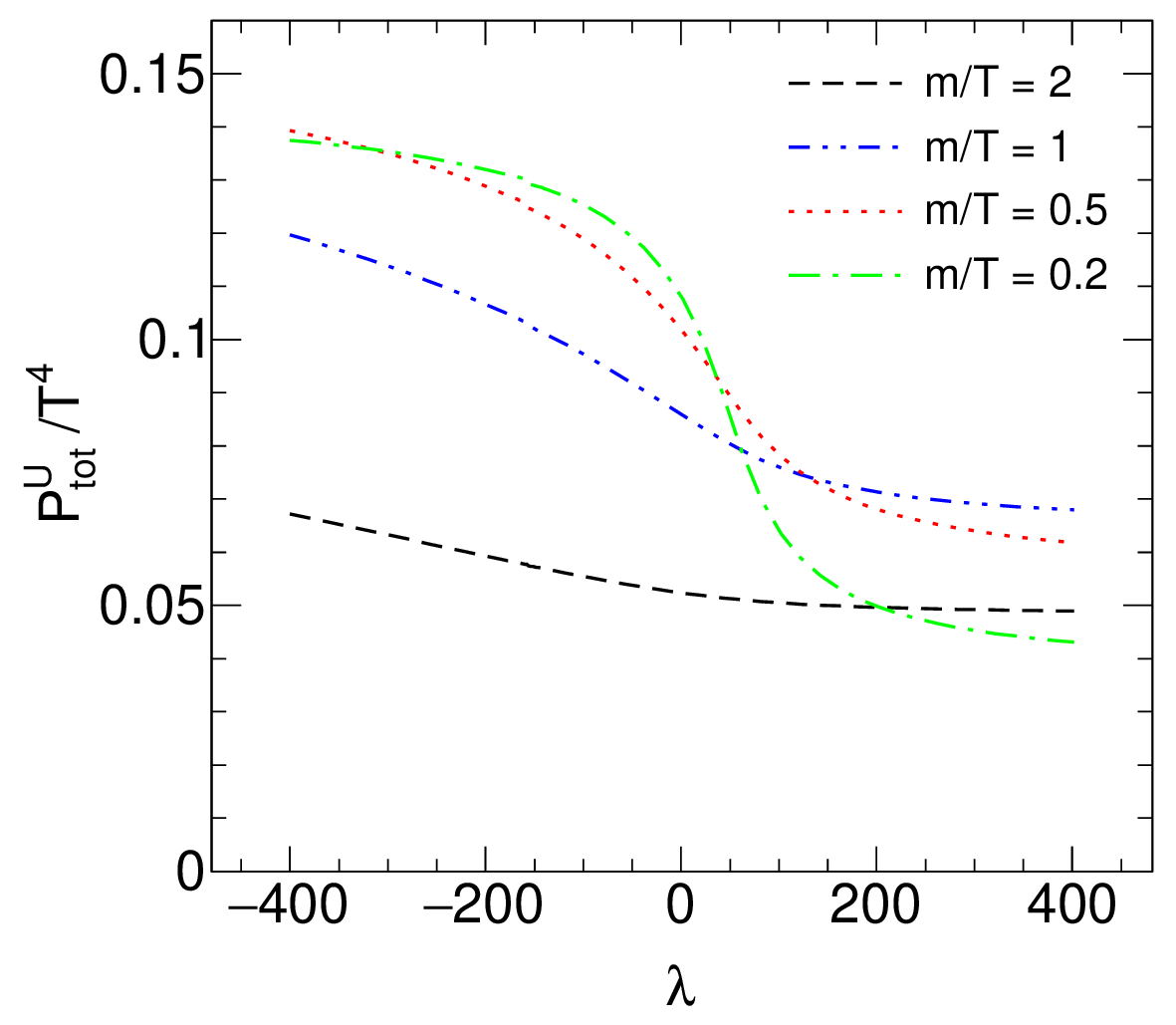}\caption{Total pressure as function of $\lambda$ for different values of the temperature.}%
\label{fig:p_lambda_total}
\end{figure}

The left panel of Fig. \ref{fig:p_lambda_uni} shows the variation of the interacting part 
of normalized pressure with $\lambda$ (excluding the contribution of the bound state) using the unitarized phase-shift.
Unlike the tree-level result (left panel of Fig. \ref{fig:P_vs_lambda_treelevel}), the
interacting pressure in the present case is discontinuous at $\lambda=\lambda_c$.
In fact, for $\lambda<\lambda_c$, the interacting part of the pressure becomes negative as a consequence of the bound state. 
The right panel of Fig.~\ref{fig:p_lambda_uni} shows the $\lambda$-dependence of interacting part of the pressure relative to that of a free gas. It shows that for $\lambda$ of the order (or larger) of $200$, the interacting part of the pressure is definitely sizable.

Figure \ref{fig:p_lambda_total}
shows the behavior of the normalized total pressure as function of $\lambda$. Here, both the contribution of the bound state and of the $\varphi\varphi$ 
interaction above threshold
are included. Quite remarkably, the total pressure is a continuous function also at $\lambda=\lambda_c$: 
the discontinuity of the interacting part of the pressure 
shown in the left panel of Fig. \ref{fig:p_lambda_total} is compensated by an analogous (but with opposite sign) jump of the bound state pressure.

\begin{figure}[ptb]
\centering
\includegraphics[width=0.48\textwidth]{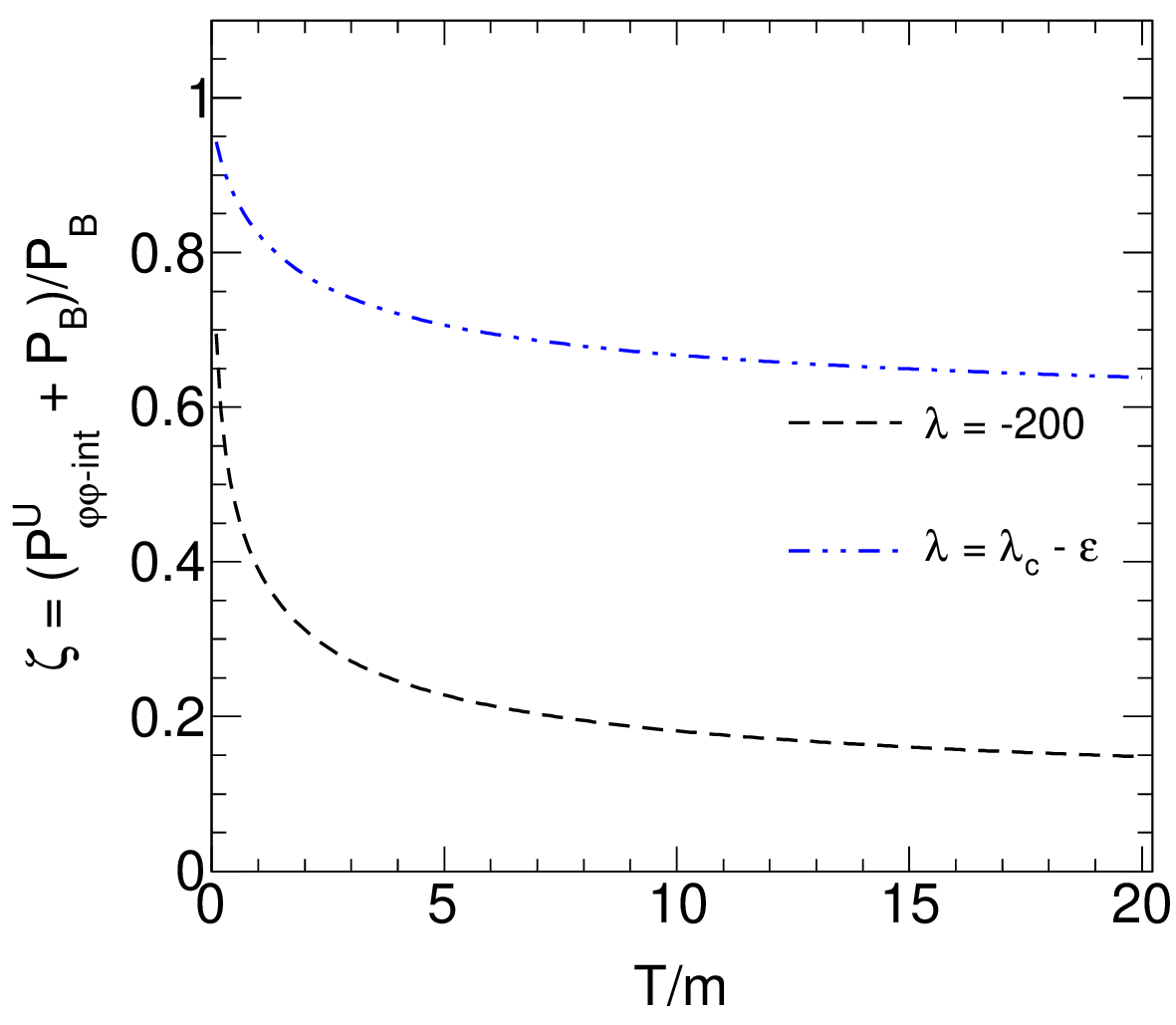}
\caption{Variation of $\zeta$ with $T/m$ for two different $\lambda$ below the critical value.}%
\label{fig:fraction_of_bs}
\end{figure}

Finally, in Fig. \ref{fig:fraction_of_bs} we show the variation of $\zeta$ 
\begin{equation}
\zeta(T,\lambda)=\frac{P_{\varphi\varphi\text{-int}}^{U}+P_{B}}{P_{B}}%
\end{equation}
as function of  $T/m$ for two different values of $\lambda$ for which the bound states forms: one just below the critical value,
$\lambda_c$, and a value sizably below it, $\lambda=-200$. This ratio approaches unity when 
$P^{U}_{\varphi\varphi-int}$ is zero.
When $\lambda$ is just below $\lambda_c$, this ratio is close to unity only
at low $T/m$; it then decreases with the increase of $T/m$ and eventually saturates around 0.6 at 
higher $T/m$.
So, even at high temperature $T/m$ this fraction is not negligible. Although the magnitude of $\zeta$ is smaller,
the trend is similar in case of $\lambda = -200$ as well. 

The results suggest that for a bound state created close to threshold (thus $\lambda$ smaller but close to $\lambda_c$), 
the bound state is indeed important, and the negative contribution to the pressure generated by the particle-particle 
interaction does not overcome the positive contribution of the bound state. In that case, one should better include the
contribution of the bound state to the pressure, but eventually one should take into account that its quantitative role 
is diminished by the interaction above threshold.

\section{Summary}

\label{sec:summary}

In this work we have investigated bound states in a thermal gas in the context
of QFT. To do this, a QFT involving a single scalar particle with mass $m$ subject to a $\varphi^{4}$-interaction
has been used. 
Besides the tree-level results, we have employed an unitarized one-loop
resummed approach for which the theory is finite and well-defined for each value
of the coupling constant $\lambda$ and for which no new energy scale appears in the theory. Moreover, for $\lambda<\lambda_c$ a bound state forms. 

The phase shift of the s-wave scattering has been calculated using the
partial wave decomposition of two body scattering and has been been used to calculate the properties of the system at finite
temperature through the phase-shift (or S-matrix) approach, according to
which the density of states is proportional to the derivative of the
phase-shift w.r.t. the running energy $\sqrt{s}$. 

For $\lambda>0$, the
contribution of the interaction to the pressure is always negative, in agreement with the repulsive nature of the interaction. On the other hand, for $\lambda_{c}<\lambda<0,$ the contribution
to the pressure is positive indicating an attractive interaction.
Below $\lambda_{c}$ the interacting part of the pressure due to two-body scattering
switches sign: it becomes negative due to the bound state below threshold. Yet, the additional contribution of a gas of bound states
makes the total pressure continuous with respect to the coupling $\lambda$. 

In summary, the contribution of the bound state to the pressure as usually calculated in thermal models is actually diminished by the contribution of the interaction among the fields, but it is not fully cancelled. Especially in the case in which the mass of the bound state is close to  $2m$ (the non-relativistic case, realized for $\lambda$ smaller but close to $\lambda_c$), the bound state has a sizable contribution to the pressure (and thus to the thermodynamics). This contribution needs to be eventually corrected by an appropriate multiplicative parameter $\zeta$ due to the role of the particle-particle interaction above threshold. Yet, it turns out to be larger than 0.6. We conclude that bound states (such as nuclei or other molecular states in QCD) should not be neglected in thermal models, even if their concrete pressure contribution can be somewhat smaller than the value of the corresponding thermal integrals. Moreover, the multiplicity of such bounds states can be calculated by the usual expression for the thermal number density, regardless of the temperature at which the gas is considered, even if it is much larger than the binding energy of the bound state. 

In the future, one can repeat the present analysis
by using other types of QFT, eventually by including fermionic fields. 
We expect that the general picture should be quite stable and independent on the precise adopted model, but it would be important to directly verify this statement. Moreover, one could also calculate the parameter $\zeta$ in some concrete examples, such as for the deuteron or for the predominantly molecular-like state $X(3872)$.

\bigskip


\textbf{Acknowledgments: }
The authors acknowledge useful discussions with W.
Broniowski and S. Mr\'{o}wczynski. SS is supported by the Polish
National Agency for Academic Exchange through Ulam Scholarship with agreement no:

PPN/ULM/2019/1/00093/U/00001. 
F. G. aknowledges acknowledges support from the
Polish National Science Centre NCN through the OPUS projects no.
2019/33/B/ST2/00613

 \appendix

\bigskip

 \bibliographystyle{jhep}

\bibliography{RefFile}

\end{document}